\documentclass[10pt,journal,final,letterpaper,twocolumn,twoside]{IEEEtran} 

\usepackage[dvips]{epsfig}
\usepackage[dvips]{graphicx}
\usepackage{amsmath,amsfonts,bm,amssymb,psfrag,ifthen,color,subfigure,amsthm}
\usepackage{algpseudocode}
\usepackage{algorithm}
\usepackage{algorithmicx}
\usepackage{setspace}
\usepackage{cite}
\usepackage{url}
\usepackage{longtable}
\usepackage{stfloats}
\usepackage{graphics,booktabs,color,subfigure}
\usepackage{float}
\usepackage{diagbox}
\usepackage{multirow}
\usepackage{array}

\allowdisplaybreaks[4]

\newcommand{\tabincell}[2]{\begin{tabular}{@{}#1@{}}#2\end{tabular}}

\begin{document}
	\title{    Covert Beamforming Design  for Integrated Radar Sensing and Communication Systems}
	\author{ Shuai~Ma,  Haihong Sheng, Ruixin Yang, Hang Li,  Youlong Wu, Chao Shen, Naofal Al-Dhahir, and Shiyin Li
		
%
 		\thanks{S. Ma is with the School of Information and Control Engineering,
 China University of Mining and Technology, Xuzhou 221116, China,
 (e-mail: mashuai001@
 cumt.edu.cn).}
%
	}
	\maketitle
	\begin{abstract}
		
		We propose covert beamforming design frameworks  for    integrated radar sensing and communication (IRSC) systems, where  the radar can   covertly communicate with legitimate users under the
		cover of the probing waveforms without being detected by the  eavesdropper.
		Specifically,   by jointly designing  the target detection
		beamformer and communication beamformer,
		we aim to maximize the radar detection mutual information (MI) (or the communication rate) subject to the covert constraint, the communication rate constraint (or the radar detection MI constraint),  and the
		total  power constraint.
		For the  perfect eavesdropper's channel state information (CSI) scenario,
		we   transform the covert beamforming design problems into a series
		of convex subproblems, by exploiting semidefinite relaxation, which can be solved via the bisection
		search method.
		Considering the high complexity of iterative optimization, we
		further propose
		a single-iterative   covert beamformer design scheme based on the zero-forcing criterion.
		For the  imperfect eavesdropper's  CSI  scenario,
		we develop a relaxation and  restriction  method to tackle the robust covert beamforming design problems.   Simulation results demonstrate    the effectiveness    of the proposed covert beamforming schemes  for
		perfect and imperfect CSI scenarios.

	\end{abstract}
	\begin{IEEEkeywords}
		Integrated Radar Sensing and Communication, Covert Beamforming Design,   Imperfect CSI.
	\end{IEEEkeywords}
	
	\IEEEpeerreviewmaketitle

	\section{Introduction}

	Compared with the separated radar and communication systems, the  integrated radar sensing  and communication (IRSC) systems enjoy a smaller platform payload and power
	consumption \cite{Liu_TC_2020}, which
	holds great potential for  both civilian and military applications,
	such as  5G vehicular network \cite{Wymeersch_WC_2017,Liu_CL_2021}, Wi-Fi based indoor
	positioning \cite{Lashkari_ICCNT_2010}, and  the advanced multi-function radio
	frequency concept (AMRFC)\cite{McCormick_RC_2017}.
	A critical challenge of developing IRSC systems is to design  integrated waveforms that can realize detection and communication simultaneously   to alleviate spectrum scarcity.
	To this end, various studies in the literature  focused on the design of dual-functional waveforms{{\cite{Hassanien_AESM_2016,Liu_TSP_2018,Li_TAES_2017,Zhou_ICCC_2018,Tong_IJSTSP_2021}}}. Due to the inherent IRSC nature of openness and broadcasting,
	the critical information embedded in the   waveform
	is susceptible
	to be intercepted and eavesdropped  by the malicious users\cite{Deligiannis_TAES_2018,Chalise_DSP_2018}.

	The security and the covertness of the radar waveforms is crucial in radar system design. Although there is extensive literature  on
	optimizing   wireless communications
	and   radar sensing simultaneously, the  information security of IRSC
	lacks in-depth research so far.
	Recently, some works \cite{Deligiannis_TAES_2018,Liu_ITWC_2018,Chalise_DSP_2018,Su_TWC_2021, Chu_ICCC_2021} have studied
	the secrecy of  IRSC   systems in terms of  the physical layer security, which focuses on
	preventing the  communication   signal   from being
	decoded by the adversary users.
	Specifically, in \cite{Deligiannis_TAES_2018},   the authors proposed three  transmit beam pattern methods  for  maximizing  the secrecy rate,
	the signal-to-interference and noise ratio (SINR), and minimizing transmit power,
	respectively.
	In \cite{Chalise_DSP_2018}, in order to reduce the risk of information security, a unified passive radar and communication system was proposed, which maximizes the signal interference-to-noise ratio (SINR) of the radar receiver under the condition that the information confidentiality rate is higher than a certain threshold.
		 In \cite{Liu_ITWC_2018},    a novel framework was proposed for the transmit beamforming of the joint RadCom system, where the beamforming schemes are designed to formulate an appropriate radar
		beampattern, while guaranteeing the SINR and power budget
		of the communication applications.
	In \cite{Su_TWC_2021}, the artificial noise (AN) was utilized  to  minimize the signal-to-noise
	ratio (SNR) of radar targets subject to the legitimate users' SINR constraint, where the target was
	regarded as a
	potential eavesdropper.
	In \cite{Chu_ICCC_2021},
	an AN-aided secure beamforming
	design  algorithm was developed  to   minimize the maximum eavesdropping SINR of the target,
	subject to the communication QoS requirements, the constant-modulus power   and
	the beampattern similarity constraints.
	Although    physical layer security technologies
	can   protect    information content from wiretapping,
	the communication behavior itself can expose sensitive information \cite{Liao_TSP_2011,Shahzad_TIFS_2021,Bash_CM_2015}.

	Different from    physical layer security technologies,
	covert communication  can guarantee   a low  probability of intercept by shielding the communication behaviors  from   potential  wardens \cite{Wang_TITS_2021,Ma_ITJ_2021,Wang_TWC_2021,Ma_TIFS_2021,Chen_TVT_2020}.
	To
	conceal the transmission from the detection by the malicious eavesdropper, one of the feasible ways  is to cover up the communication signal with the radar signal.
	In  \cite{Blunt_IWDDC_2007,Blunt_ICEAA_2007}, the authors  first  hide  the communication symbol
	by utilizing the  multi-path effects of tag/transponder (simply denoted as the "tag").
	Specifically,  by using
	tags  to modulate the reflection of the
	incident radar waveform into  communication waveforms,
	the modulated communication
	waveform is   embedded into the ambient radar pulse
	scattering, which acts as masking interference to maintain a low intercept probability.
	Note that, the covert communications in \cite{Blunt_IWDDC_2007,Blunt_ICEAA_2007} depend on extra environmental conditions (the multi-path effects of tags), which may not always be available  in practice.

	So far, covert communication for IRSC systems has not been well investigated. Particularly, the covert design framework for the IRSC systems has not been rigorously established. Against this background,   this paper  establishes a covert communication optimization framework  for   IRSC systems{\footnote{We focus on beamforming optimization for single antenna receivers \cite{Chalise_DSP_2018}.
	}}. To be specific, under the covert constraint, we propose two covert beamforming optimization frameworks for mutual information and covert rate maximization.
	The frameworks consider both the cases of perfect and imperfect channel state information. The  main contributions of this work are summarized follows:
	
	\begin{itemize}
		
		\item  Considering  Willie's (eavesdropper) channel state information
		(WCSI) being available at the radar, both  the radar detection  mutual information (MI) maximization and covert rate maximization  are studied under both the covert   and total transmit  power constraint, which are non-convex and hard to solve.
		Based on   semidefinite relaxation (SDR), we first relax the covert beamforming design optimization   problem   into a series of
		convex subproblems, and then  efficiently solved them via the bisection search method.

		\item   In order to avoid the high complexity of iterative optimization, we
		further propose
		a single-iterative   covert beamformer design scheme based on the zero-forcing criterion. Specifically,
		by   designing   the target detection
		beamformer as a cover,   we optimize the communication beamformer to be orthogonal to the WCSI, and thus the communication signals are projected onto the null space of  Willie's channel, and
		achieve   covert communications.

		\item Furthermore, with imperfect WCSI, we develop a relaxation and  restriction method to tackle the robust covert beamforming design optimization problems.
		Specifically, by exploiting   the piecewise monotonicity property  of the covert function, we
		first transform the covert  constraint  into a simplified and equivalent form   facilitating
		robust beamforming  design. Then, we relax the robust covert beamforming design optimization problems
		based on SDR,  and restrict it into a convex    semidefinite program (SDP). Extensive numerical results quantify the effects of the key design parameters on the system performance.
		
	\end{itemize}

	The remaining part of this paper is organized as follows. In
	Section II, we describe    the IRSC system model and problem formulation.
	Then, we present our covert beamforming design with perfect WCSI in Section III. In Section IV, the robust covert beamforming design is developed for imperfect WCSI. Section V presents numerical results of the proposed covert beamforming design frameworks, and conclusions are drawn in
	Section VI. Table I presents the means of the
	key notations in  this paper.
	\begin{table}[htbp]
		\caption{Summary of Key Notations}
		\label{tablepar}
		\centering
		\begin{tabular}{|m{1.8cm}<{\centering}|m{6.2cm}|}
			\hline
			\rule{0pt}{8pt}Notation  &    Description \\ \hline
			\rule{0pt}{7.5pt}$\rm{s_{\rm{R}}}$ &  \tabincell{c}{Detection signal} \\ \hline
			\rule{0pt}{7.5pt}$\rm{s_{\rm{C}}}$ &  \tabincell{c}{Communication signal}\\ \hline
			\rule{0pt}{7.5pt}${{\bf{h}}_{\rm{T}}} $ &  \tabincell{c}{Channel gain vector of radar-target path}\\ \hline
			\rule{0pt}{7.5pt}${{\bf{h}}_{\rm{B}}}$ &  \tabincell{c}{Channel gain vector of radar-Bob path}\\ \hline		\rule{0pt}{7.5pt}${{\bf{h}}_{\rm{W}}}$ &  \tabincell{c}{Channel gain vector of radar-Willie path}\\ \hline
			\rule{0pt}{7.5pt}${{\bf{w}}_{\rm{R,0}}}$ &  \tabincell{c}{Beamformer  for ${s_{\rm{R}} }$}\\ \hline
			\rule{0pt}{7.5pt}${{\bf{w}}_{\rm{R,1}}}$ &  \tabincell{c}{Beamformer   for $\rm{s_{\rm{R}}}$}\\ \hline
			\rule{0pt}{7.5pt} ${{\cal H}_0}$ &  The null hypothesis that the radar only sends $\rm{s_{\rm{R}}}$ \\ \hline
			\rule{0pt}{7.5pt} ${{\cal H}_1}$  &  The hypothesis that the radar  sends both $\rm{s_{\rm{R}}}$ and $\rm{s_{\rm{C}}}$  \\ \hline
			\rule{0pt}{7.5pt} ${y_{\rm{W}}}$  &  \tabincell{c}{Received signal of Willie}  \\ \hline
			\rule{0pt}{7.5pt} ${p_0}\left( {{y_{\rm{W}}}} \right)$  &  \tabincell{c}{Likelihood function of ${{y_{\rm{W}}}}$ under ${{\cal H}_0}$}  \\ \hline
			\rule{0pt}{7.5pt} ${p_1}\left( {{y_{\rm{W}}}} \right) $   &  \tabincell{c}{Likelihood function of ${{y_{\rm{W}}}}$ under  ${{\cal H}_1}$}  \\ \hline
			\rule{0pt}{7.5pt} $D\left( {{p_1}\left\| {{p_0}} \right.} \right)$ &   Kullback-Leibler (KL) divergence from $p_1(y_{\rm{W}})$ to  $p_0(y_{\rm{W}})$ \\ \hline
			\rule{0pt}{7.5pt}$\xi $ &  Total detection error probability  \\ \hline
			\rule{0pt}{7.5pt}$ P \left( {{{\cal D}_1}\left| {{{\cal H}_0}} \right.} \right)$ &  False alarm (FA) probability  \\ \hline
			\rule{0pt}{7.5pt}$P \left( {{{\cal D}_0}\left| {{{\cal H}_1}} \right.} \right)$ &  Missed detection (MD) probability  \\ \hline		
		\end{tabular}
	\end{table}

	
	\section{System Model and Problem Formulation }
	\begin{figure}[htbp]
		\centering
		\includegraphics[height=4.5cm]{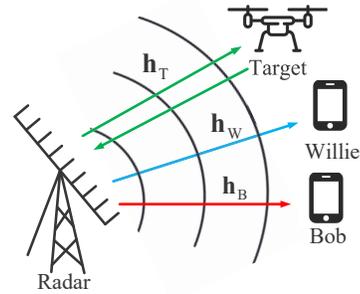}
		\caption{~The {schematic diagram} of a integrated radar-communication system.}
		\label{img1}
	\end{figure}
	
	Considering an integrated radar-communication system,  as shown in Fig. $\ref{img1}$,
which includes a radar, a target, and a legitimate receiver (Bob).
The radar  with    ${N}$ antennas   is capable of    detecting target and transmitting signals to the Bob simultaneously. While Bob is installed with a single antenna, and thus the reflected signals   from
  the target  can be ignored at Bob.
 In addition, an eavesdropper (Willie) with a single antenna keeps detecting the communication signals between
 the radar and Bob, and tries to identify whether the radar is transmitting information to Bob.
    Moreover,
 ${{\bf{h}}_{\rm{T}}} \in {\mathbb{C}^{N \times 1}}$, ${{\bf{h}}_{\rm{B}}} \in {\mathbb{C}^{N \times 1}}$ and  ${{\bf{h}}_{\rm{W}}} \in {\mathbb{C}^{N \times 1}}$  denote channel state information (CSI) of the Radar-Target path, the Radar-Bob path and  the Radar-Willie path, respectively.
      All channels  are assumed followed the Rayleigh flat fading model  \cite{Shahzad_VTC_2017}, i.e., ${{\bf{h}}_{{\rm{B}}}} \sim \;{\cal C}{\cal N}\left( {{\bf{0}},\sigma _{{\rm{1}}}^{\rm{2}}{\bf{I}}} \right),{{\bf{h}}_{{\rm{W}}}} \sim \;{\cal C}{\cal N}\left( {{\bf{0}},\sigma _{{\rm{2}}}^2{\bf{I}}} \right)$,
	where $\sigma _{{\rm{1}}}^2$ and $\sigma _{{\rm{2}}}^2$ denote the variances of  channels ${{\bf{h}}_{{\rm{T}}}}$, ${{\bf{h}}_{{\rm{B}}}}$, and  ${{\bf{h}}_{{\rm{W}}}}$, respectively.
  Let $\rm{s_{\rm{R}}}$ and $\rm{s_{\rm{C}}}$ respectively  denote the detection signal  and    and the communication signal of the radar {\footnote{In   typical scenarios, the power of the detection signal  is higher than  that of the communication signal.
	}}. Without loss of generality, we assume that ${{\mathbb E}}\left\{ {{{\left| {{s_{\rm{R}}}} \right|}^2}} \right\} = 1$, ${{\mathbb E}}\left\{ {{{\left| {{s_{\rm{C}}}} \right|}^2}} \right\} = 1$.
%
	
	\subsection{Signal Model}
	\subsubsection{Detection Only}

		The Radar-Target path	${\bf h}_{\rm{T}}$ can be expressed as
		\begin{align}
			{{\bf{h}}_{\rm{T}}} \triangleq {{\bf{a}}_T}(\theta ),
		\end{align}
	where $\theta$ represents the azimuth angle of the target, and ${{\bf{a}}_T}(\theta )$ represents  the transmit steering vector.	
	 Meanwhile, assume that $\mathbf{a}_{T}(\theta) \triangleq \mathbf{a}_{R}(\theta)$,
where ${{\bf{a}}_R}(\theta )$ represents  the receive steering vector \cite{Khawar_ISJ_2015}. Therefore, when the radar sends only detection signals and not the communication signals, the reflected signals from the target can be written as
	\begin{align}\label{part1_03_01}
		{{\bf{y}}_{\rm{R}}} = \alpha {{\bf{h}}_{\rm{T}}}{\bf{h}}_{\rm{T}}^H{{\bf{w}}_{{\rm{R}},{\rm{0}}}}{s_{\rm{R}}} + {{\bf{z}}_{\rm{R}}},
	\end{align}	
	where $\alpha$ denotes the path-loss coefficient, ${{\bf{w}}_{\rm{R,0}}}$ denotes the radar transmission beamformer vector for ${s_{\rm{R}} }$ when the radar only sends detection signals, and ${\bf{z}_{\rm{R}}}\sim {\cal CN}\left({{\rm{0}},\sigma _{\rm{R}}^2{\bf{I}}} \right)$ is the received noise at the radar.
			
	Here, we rely on the following standard assumptions:
 parameters $\alpha$ and $\theta$ can be estimated from the received signal \cite{chalise_ISPL_2017}.
	Since the transmitted signals can be omni-directional, the legitimate user Bob could receive the detection signals as well. For Bob, we have
	\begin{align}\label{part1_01_1}
		{y_{\rm{B}}} = {\bf{h}}_{\rm{B}}^H{{\bf{w}}_{{\rm{R}},{\rm{0}}}}{s_{\rm{R}}} + {z_{\rm{B}}},
	\end{align}
	where ${z_{\rm{B}}}\sim {\cal CN}\left( {{\rm{0}},\sigma _{\rm{B}}^2} \right)$ is the receiver noise of Bob.

	\subsubsection{Detection and Communication}	
	The communication signals from the radar are also omni-directional. Moreover, since the radar has multiple antennas, the reflected communication signals from the target can also be captured. Thus, when the detection and the communication functions are performed simultaneously, the received signal at the radar is
	\begin{align}\label{part1_03_2}
		{{\bf{y}}_{\rm{R}}} = \alpha {{\bf{h}}_{\rm{T}}}{\bf{h}}_{\rm{T}}^H{{\bf{w}}_{{\rm{R}},0}}{s_{\rm{R}}} + \alpha {{\bf{h}}_{\rm{T}}}{\bf{h}}_{\rm{T}}^H{{\bf{w}}_{{\rm{R}},{\rm{1}}}}{s_{\rm{C}}} + {{\bf{z}}_{\rm{R}}},
	\end{align}
	where ${{\bf{w}}_{\rm{R,1}}}$ denotes the beamformer vectors for ${s_{\rm{C}} }$ when the radar sends detection signals and communication signals.
	For the legitimate user Bob, the received signal can be written   as	
	\begin{align}\label{part1_01_2}
		{y_{\rm{B}}} = {{\bf{h}}_{\rm{B}}^H{{\bf{w}}_{{\rm{R}},0}}{s_{\rm{R}}} + {\bf{h}}_{\rm{B}}^H{{\bf{w}}_{{\rm{R}},{\rm{1}}}}{s_{\rm{C}}} + {z_{\rm{B}}}}.
	\end{align}
	%
	
	\subsection{Performance Metrics}
	The considered system requires several metrics to measure the performance, according to the operating mode.
	
	\subsubsection{Detection Only}
	
	For radar, by receiving the echo from the target, the radar estimates the channel ${\bf{h}}_{\rm{T}}$ and identifies some unknown characteristics about the target. Such characteristics can be quantified by using a tool from information theory which uses the observation at the channel output and reduces the uncertainty of prior information via certain code design. This tool successfully defines the information transmission capability of the communication channel \cite{Bell_TIT_1993}. This idea can also be used for the radar transmission design \cite{Bell_TIT_1993, Tang_TSP_2019, Bica_ICASSP_2016}.
	
	Specifically, after receiving ${{\bf{y}}_{\rm{R}}}$, the priori uncertainty of the target decreases since there is some information about the target contained in ${\bf{h}}_{\rm{T}}$ \cite{Bell_TIT_1993}. Thus,
 we adopt the mutual information (MI) between ${\bf{h}}_{\rm{T}}$ and ${{\bf{y}}_{\rm{R}}}$ given the transmission signal ${s_{\rm{R}}}$, i.e.,
  $	{\rm{I}}\left( {{{\bf{y}}_{\rm{R}}};{{\bf{h}}_{\rm{T}}}\left| {{s_{\rm{R}}}} \right.} \right){\rm{ }}$, to characterize  how much information the radar can learn from ${{\bf{y}}_{\rm{R}}}$,
   which is given as
	\begin{align}\label{MI_01}
		\begin{split}
			{\rm{I}}\left( {{{\bf{y}}_{\rm{R}}};{{\bf{h}}_{\rm{T}}}\left| {{s_{\rm{R}}}} \right.} \right){\rm{ }}
			&= \frac{1}{2}\log \left( {1 + \frac{{{{\left| \alpha  \right|}^2}{{\left| {{\bf{h}}_{\rm{T}}^H{{\bf{w}}_{{\rm{R}},{\rm{0}}}}} \right|}^2}{{\left\| {{{\bf{h}}_{\rm{T}}}} \right\|}^2}}}{{\sigma _{\rm{R}}^2}}} \right).
		\end{split}
	\end{align}
	
	\subsubsection{Detection and Communication}
	For the MI at the radar, according to   $\eqref{part1_03_2}$,
	we have{\footnote{{{{For the radar and communication receiver, the detection signal $\rm{s_{\rm{R}}}$ is deterministic to facilitate receiver signal extraction.
	}}}}}
	\begin{subequations}\label{MI}
		\begin{align}
			\rm{I}\left( {{{\bf{y}}_{\rm{R}}};{{\bf{h}}_{\rm{T}}}\left| {{s_{\rm{R}}}} \right.} \right)& = h\left( {{{\bf{y}}_{\rm{R}}}\left| {{s_{\rm{R}}}} \right.} \right) - h\left( {{{\bf{y}}_{\rm{R}}}\left| {{{\bf{h}}_{\rm{T}}},{s_{\rm{R}}}} \right.} \right)\nonumber\\
			& = \frac{1}{2}\log \left( {1 + \frac{{{{\left| \alpha  \right|}^2}{{\left| {{\bf{h}}_{\rm{T}}^H{{\bf{w}}_{{\rm{R,0}}}}} \right|}^2}{{\left\| {{{\bf{h}}_{\rm{T}}}} \right\|}^2}}}{{{{\left| \alpha  \right|}^2}{{\left| {{\bf{h}}_{\rm{T}}^H{{\bf{w}}_{{\rm{R,1}}}}} \right|}^2}{{\left\| {{{\bf{h}}_{\rm{T}}}} \right\|}^2} + \sigma _{{{\bf{z}}_{\rm{R}}}}^2}}} \right).\nonumber
		\end{align}
	\end{subequations}
	
 	 Thus, based on $\eqref{part1_01_2}$,   the achievable rate of Bob  can be expressed as \cite{Yan_TIFS_2019}	
	\begin{align}\label{SINRB}
		{R_{\rm{B}}}\left( {{{\bf{w}}_{{\rm{R}},{\rm{0}}}},{{\bf{w}}_{{\rm{R}},{\rm{1}}}}} \right) = {\log _2}\left( {1 + \frac{{{{\left| {{\bf{h}}_{\rm{B}}^H{{\bf{w}}_{{\rm{R,1}}}}} \right|}^2}}}{{{{\left| {{\bf{h}}_{\rm{B}}^H{{\bf{w}}_{{\rm{R,0}}}}} \right|}^2} + \sigma _{\rm{B}}^2}}} \right).
	\end{align}

	\subsection{Covert Constraints} In the considered system, the eavesdropper Willie aims to determine whether Bob is communicating with the radar. Mathematically, Willie needs to perform a hypothesis test between the two hypotheses from its received signal ${y_{\rm{W}}}$. Here, ${{\cal H}_0}$ represents the null hypothesis, i.e., the radar only sends detection signals; ${{\cal H}_1}$ represents the other hypothesis, where the radar sends detection signals and communication signals. Thus, ${y_{\rm{W}}}$ can be written as
	\begin{align}\label{part1_02}
		{y_{\rm{W}}} = \left\{ {\begin{array}{*{20}{c}}
				{{\bf{h}}_{\rm{W}}^H{{\bf{w}}_{{\rm{R}},{\rm{0}}}}{s_{\rm{R}}} + {z_{\rm{W}}},}&{{{\cal H}_0}},\\
				{{\bf{h}}_{\rm{W}}^H{{\bf{w}}_{{\rm{R}},0}}{s_{\rm{R}}} + {\bf{h}}_{\rm{W}}^H{{\bf{w}}_{{\rm{R,1}}}}{s_{\rm{C}}} + {z_{\rm{W}}},}&{{{\cal H}_1}},
		\end{array}} \right.
	\end{align}
	where ${z_{\rm{W}}}\sim {\cal CN}\left( {{\rm{0}},\sigma _{\rm{W}}^2} \right)$  is the receiving noise at Willie.  Note that, when the radar performs tracking and communication, the radar needs to assume the existence of Willie and tries to conceal the communication with Bob.

In the following, we will introduce the covert constraint  to describe the level of covertness.
	  Specifically, for Willie,  let ${p_0}\left( {{y_{\rm{W}}}} \right)$ and ${p_1}\left( {{y_{\rm{W}}}} \right) $ to represent the likelihood function of ${{y_{\rm{W}}}}$ under ${{\cal H}_0}$ and ${{\cal H}_1}$, respectively. According to $\eqref{part1_02}$, ${p_0}\left( {{y_{\rm{W}}}} \right)$ and ${p_1}\left( {{y_{\rm{W}}}} \right) $ can be expressed as
	\begin{subequations}\label{part1_05}
		\begin{align}
			& {p_0}\left( {{y_{\rm{W}}}} \right) = \frac{1}{{\pi {\lambda _0}}}\exp \left( { - \frac{{{{\left| {{y_{\rm{W}}}} \right|}^2}}}{{{\lambda _0}}}} \right), \\
			&{p_1}\left( {{y_{\rm{W}}}} \right) = \frac{1}{{\pi {\lambda _1}}}\exp \left( { - \frac{{{{\left| {{y_{\rm{W}}}} \right|}^2}}}{{{\lambda _1}}}} \right),
		\end{align}
	\end{subequations}
	where ${\lambda _{\rm{0}}} \buildrel \Delta \over = {\left| {{\bf{h}}_{\rm{W}}^H{{\bf{w}}_{\rm{R,0}}}} \right|^2} + \sigma _{\rm{W}}^2$ and $
	{\lambda _{\rm{1}}} \buildrel \Delta \over = {\left| {{\bf{h}}_{\rm{W}}^H{{\bf{w}}_{\rm{R,0}}}} \right|^2} + {\left| {{\bf{h}}_{{\rm{W}}}^H{{\bf{w}}_{\rm{R,1}}}} \right|^2} + \sigma _{\rm{W}}^2$.
		Moreover,  let $D\left( {{p_0}\left\| {{p_1}} \right.} \right)$   denote the  KL divergence from $p_0(y_{\rm{W}})$ to  $p_1(y_{\rm{W}})$, and
	$D\left( {{p_1}\left\| {{p_0}} \right.} \right)$ denote the KL divergence from $p_1(y_{\rm{W}})$ to  $p_0(y_{\rm{W}})$.
	According to $\eqref{part1_05}$,
	$D\left( {{p_0}\left\| {{p_1}} \right.} \right)$   and
	$D\left( {{p_1}\left\| {{p_0}} \right.} \right)$ are  respectively  given as
	\begin{subequations}\label{part1_08}
		\begin{align}
			D\left( {{p_0}\left\| {{p_1}} \right.} \right) &= \int_{ - \infty }^{ + \infty } {{p_0}\left( {{y_{\text{W}}}} \right)\ln \frac{{{p_0}\left( {{y_{\text{W}}}} \right)}}{{{p_1}\left( {{y_{\text{W}}}} \right)}}} dy  = \ln \frac{{{\lambda _1}}}{{{\lambda _0}}} + \frac{{{\lambda _0}}}{{{\lambda _1}}} - 1,\\
			D\left( {{p_1}\left\| {{p_0}} \right.} \right) &= \int_{ - \infty }^{ + \infty } {{p_1}\left( {{y_{\text{W}}}} \right)\ln \frac{{{p_1}\left( {{y_{\text{W}}}} \right)}}{{{p_0}\left( {{y_{\text{W}}}} \right)}}} dy = \ln \frac{{{\lambda _0}}}{{{\lambda _1}}} + \frac{{{\lambda _1}}}{{{\lambda _0}}} - 1.
		\end{align}
	\end{subequations}
	
	\subsubsection{Detection by Willie}
	Let ${{{\cal D}_1}}$ and ${{{\cal D}_0}}$ denote the  binary decisions that correspond to hypotheses ${{{\cal H}_0}}$ and  ${{{\cal H}_1}}$, respectively.
	The detecting performance at Willie is described by the FA probability $ P \left( {{{\cal D}_1}\left| {{{\cal H}_0}} \right.} \right)$ and the MD probability $P \left( {{{\cal D}_0}\left| {{{\cal H}_1}} \right.} \right)$.	The total detection error probability $\xi $ can be used to measure the covertness of the system \cite{Lehmann_Springer_2005}, i.e.,
	\begin{align}\label{part1_09}
		\xi = P \left( {{{\cal D}_1}\left| {{{\cal H}_0}} \right.} \right)+P \left( {{{\cal D}_0}\left| {{{\cal H}_1}} \right.} \right).
	\end{align}
	
	To further evaluate $\eqref{part1_09}$, we assume that the probabilities ${{\cal H}_0}$ and ${{\cal H}_1}$ are equal. By applying the Neyman-Pearson criterion \cite{Lehmann_Springer_2005}, the optimal  rule  for Willie to
	minimize $\xi$ as the   likelihood
	ratio test, i.e.,
	\begin{align}\label{C_17}
		\frac{{{p_1}\left( {{y_{\rm{W}}}} \right)}}{{{p_0}\left( {{y_{\rm{W}}}} \right)}}\frac{{\mathop  > \limits^{{D_1}} }}{{\mathop  < \limits_{{D_0}} }}1.
	\end{align}
	
	After some algebraic operations, $\eqref{C_17}$ can be equivalently reformulated as
	\begin{align}\label{C_17_1}
		{\left| {{y_{\rm{W}}}} \right|^2} \frac{{\mathop  > \limits^{{{\cal D}_1}} }}{{\mathop  < \limits_{{{\cal D}_0}} }} {\phi ^*},
	\end{align}
	where  ${\left| {{y_{\rm{W}}}} \right|^2}$ is the average power at Willie, and $\phi ^*$ denotes the optimal detection threshold of Willie, which is given by
	\begin{align}\label{C_17_2}
		{\phi ^*} \buildrel \Delta \over = \frac{{{\lambda _0}{\lambda _1}}}{{{\lambda _1} - {\lambda _0}}}{\rm{ln}}\frac{{{\lambda _1}}}{{{\lambda _0}}}.
	\end{align}
	
	According to $\eqref{part1_05}$,   the cumulative density functions (CDFs) of ${\left| {{y_{\rm{W}}}} \right|^2}$ under ${{{\cal H}_{\rm{0}}}}$ and ${{{\cal H}_{\rm{1}}}}$ are given by
	\begin{subequations}\label{C_9}
		\begin{align}
			&\Pr \left( {\left. {{{\left| {{y_{\rm{W}}}} \right|}^2}} \right|{\cal{H}_{\rm{0}}}} \right) = 1 - \exp \left( { - \frac{{{{\left| {{y_{\rm{W}}}} \right|}^2}}}{{{\lambda _0}}}} \right), \\
			&\Pr \left( {\left. {{{\left| {{y_{\rm{W}}}} \right|}^2}} \right|{\cal{H}_{\rm{1}}}} \right) = 1 - \exp \left( { - \frac{{{{\left| {{y_{\rm{W}}}} \right|}^2}}}{{{\lambda _1}}}} \right).
		\end{align}
	\end{subequations}
	
	Based on the optimal detection threshold ${\phi ^*}$, we obtain
	\begin{subequations}
		\begin{align}
			& P \left( {{{\cal D}_1}\left| {{{\cal H}_0}} \right.} \right)= {\Pr}\left( {{{\left| {{y_{\rm{W}}}} \right|}^2} \ge {\phi ^*}|{{{\cal H}_{\rm{0}}}}} \right) = {\left( {\frac{{{\lambda _1}}}{{{\lambda _0}}}} \right)^{ - \frac{{{\lambda _1}}}{{{\lambda _1} - {\lambda _0}}}}},    \nonumber   \\
			& P \left( {{{\cal D}_0}\left| {{{\cal H}_1}} \right.} \right) = {\Pr}\left( {{{\left| {{y_{\rm{W}}}} \right|}^2} \le {\phi ^*}|{{{\cal H}_{\rm{1}}}}} \right)  = 1 - {\left( {\frac{{{\lambda _1}}}{{{\lambda _0}}}} \right)^{ - \frac{{{\lambda _0}}}{{{\lambda _1} - {\lambda _0}}}}}.\nonumber
		\end{align}
	\end{subequations}
	
	\subsubsection{Definition of Covert Constraint}
	The eavesdropper Willie tries to find the best detector with the least total detection error probability $\xi^*$ \cite{Yan_TWC_2019}. From the legitimate user perspective, an effective covert communication should guarantee that no matter what strategy Willie adopts, for any given small constant $\varepsilon  \in \left[ {0,1} \right]$, the following criterion can be satisfied 	
	$\xi^*  \ge 1 - \varepsilon \label{part1_06} $\cite{Lehmann_Springer_2005}.
	Thus, when Willie adopts the optimal detector, we have \cite{Lehmann_Springer_2005, Yan_TWC_2019, Bash_JSAC_2013, Ben-Naim_FTE_2014}
	\begin{align}\label{V_T_1}
		\xi^* =  1 - {V_{\rm{T}}}\left( {{p_0},{p_1}} \right).
	\end{align}
	where ${V_{\rm{T}}}\left( {{p_0},{p_1}} \right)$ is expressed as the total variation distance between ${p_0}$ and ${p_1}$.
	
	Furthermore, based on Pinsker's inequality \cite{Cover_book_1999}, we obtain
	\begin{subequations}\label{V_T_2}
		\begin{align}
			& { V_{\rm{T}}}\left( {{p_0},{p_1}} \right) \le \sqrt {\frac{1}{2}D\left( {{p_0}\left\| {{p_1}} \right.} \right)}, \\
			&{ V_{\rm{T}}}\left( {{p_0},{p_1}} \right) \le \sqrt {\frac{1}{2}D\left( {{p_1}\left\| {{p_0}} \right.} \right)}.
		\end{align}
	\end{subequations}
	
	Therefore, by combining the   formulas  \eqref{V_T_1} and \eqref{V_T_2}, we obtain the tractable covert constraints as follows
	\begin{subequations}\label{part1_07}
		\begin{align}
			D\left( {{p_0}\left\| {{p_1}} \right.} \right) \le 2{\varepsilon ^2},\label{part1_07_1}\\
			D\left( {{p_1}\left\| {{p_0}} \right.} \right) \le 2{\varepsilon ^2}{{,}}\label{part1_07_2}
		\end{align}
	\end{subequations}
	which are determined by the transmission variables and the receiver noise (see $\eqref{part1_08}$).
	
	\subsection{Problem Formulation}
	Given the two performance metrics when the radar performs the detection and communication at the same time, an intuitive optimization goal is to make these two metrics as large as possible.
	Thus, we obtain the following two terms optimization problems, i.e., the  radar MI maximization problem and the communication rate maximization problem.
	\subsubsection{MI Maximization}
	As   described above, a larger MI indicates more possibilities to identify the information about the target from the received signals. Thus, we aim to design beamforming vectors with the maximum MI $\eqref{MI}$ while satisfying the covert constraint, the covert rate requirement  and the total transmit  power constraints, which can be mathematically formulated as
	\begin{subequations}\label{C2}
		\begin{align}
			\mathop {\max }\limits_{{{\bf{w}}_{{\rm{R,0}}}},{{\bf{w}}_{{\rm{R,1}}}}}& {\rm{I}}\left( {{\bf{y}_{\rm{R}}};{{\bf{h}}_{\rm{T}}}\left| {{s_{\rm{R}}}} \right.} \right)\\
			{\rm{s}}.{\rm{t}}. \quad
			&{R_{\rm{B}}}\left( {{{\bf{w}}_{{\rm{R}},{\rm{0}}}},{{\bf{w}}_{{\rm{R}},{\rm{1}}}}} \right) \geq \beta ,\label{C2-R}\\
			&{\left\| {{{\bf{w}}_{{\rm{R}},{\rm{0}}}}} \right\|^2} + {\left\| {{{\bf{w}}_{{\rm{R}},{\rm{1}}}}} \right\|^2} \le {P_{{\rm{total}}}},\label{C2-P}\\
			&\eqref{part1_07_1} {\quad \rm{or} \quad} \eqref{part1_07_2},\nonumber
		\end{align}
	\end{subequations}
where    ${P_{{\rm{total}}}}$ denotes the maximum radar transmit power. 

	\subsubsection{Rate Maximization}
	The objective of this problem is to design a beamformer that maximizes the rate achieved by Bob while satisfying the required covertness constraints, the MI requirement, and the total transmit power constraints.
	Mathematically, the problem of maximizing the achievable rates can be expressed as
	\begin{subequations}\label{C22}
		\begin{align}
			\mathop {\max }\limits_{{{\bf{w}}_{{\rm{R,0}}}},{{\bf{w}}_{{\rm{R,1}}}}}& {R_{\rm{B}}}\left( {{{\bf{w}}_{{\rm{R}},{\rm{0}}}},{{\bf{w}}_{{\rm{R}},{\rm{1}}}}} \right)\\
			{\rm{s}}.{\rm{t}}. \quad
			&{\rm{I}}\left( {{\bf{y}_{\rm{R}}};{{\bf{h}}_{\rm{T}}}\left| {{s_{\rm{R}}}} \right.} \right) \geq \gamma ,\label{R-R}\\
			&{\left\| {{{\bf{w}}_{{\rm{R}},{\rm{0}}}}} \right\|^2} + {\left\| {{{\bf{w}}_{{\rm{R}},{\rm{1}}}}} \right\|^2} \le {P_{{\rm{total}}}},\label{R-P}\\
			&\eqref{part1_07_1} {\quad \rm{or} \quad} \eqref{part1_07_2}.\nonumber
		\end{align}
	\end{subequations}
	\textbf{Remarks: }Note that, when the radar performs detection only, the corresponding optimization problem becomes  a simplified version of problem $\eqref{C2}$. Thus, we omit this case, and only investigate the  integrated
radar sensing and communication case.
		From the above formulations,   the two optimization problems \eqref{C2} and \eqref{C22} have similar structure. Basically, for the MI and the covert rate, the beamformer splits the available power to maximize one term while satisfying the requirement of the other term. Hence, there should exist a tradeoff between the MI and the covert rate. In the following, we will focus on the investigation of the MI maximization problem, and numerically discuss the tradeoff.
	
	\section{Beamforming Design based on PERFECT WCSI}
	
	We  first consider a typical scenario, where Willie
	is a legitimate user. In this case, the radar may obtain the full CSI of ${{\bf{h}}_{\rm{W}}}$, and uses it to help Bob to hide from Willie under ${{\cal H}_1}$ \cite{Bash_JSAC_2013,Yan_TIFS_2019}. Then, the covert constraint becomes
	\begin{subequations}\label{D}	
		\begin{align}
			D\left( {{p_0}\left\| {{p_1}} \right.} \right) = 0,\\
			D\left( {{p_1}\left\| {{p_0}} \right.} \right) = 0.
		\end{align}
	\end{subequations}
	
	Note that, $\eqref{D}$ also implies the perfect covert transmission.
	
	
	\subsection{MI Maximization}
	
	After applying formula $\eqref{C2}$, the MI maximization problem can be reformulated as
	\begin{subequations}\label{part22_02}
		\begin{align}
			\mathop {\max }\limits_{{{\bf{w}}_{{\rm{R,0}}}},{{\bf{w}}_{{\rm{R,1}}}}}&  \frac{1}{2}\log \left( {1 + \frac{{{\left| \alpha  \right|^2}{{\left| {{\bf{h}}_{\rm{T}}^H{{\bf{w}}_{{\rm{R,0}}}}} \right|}^2}{{\left\| {{{\bf{h}}_{\rm{T}}}} \right\|}^2}}}{{{\left| \alpha  \right|^2}{{\left| {{\bf{h}}_{\rm{T}}^H{{\bf{w}}_{{\rm{R,1}}}}} \right|}^2}{{\left\| {{{\bf{h}}_{\rm{T}}}} \right\|}^2} + \sigma _{\rm{R}}^2}}} \right) \\
			{\rm{s}}.{\rm{t}}. \quad	&
			{\left| {{\bf{h}}_{\rm{W}}^H{{\bf{w}}_{{\rm{R,1}}}}} \right|^2} = 0,\\
			& \frac{{{{\left| {{\bf{h}}_{\rm{B}}^H{{\bf{w}}_{{\rm{R,1}}}}} \right|}^2}}}{{{{\left| {{\bf{h}}_{\rm{B}}^H{{\bf{w}}_{{\rm{R,0}}}}} \right|}^2} + \sigma _{\rm{B}}^2}} \geq \beta ,\\
			&{\left\| {{{\bf{w}}_{{\rm{R}},{\rm{0}}}}} \right\|^2} + {\left\| {{{\bf{w}}_{{\rm{R}},{\rm{1}}}}} \right\|^2} \le {P_{{\rm{total}}}}.
		\end{align}
	\end{subequations}
	
	Next, we propose two different methods, namely covert beamformer and zero-forcing beamformer, to solve   problem \eqref{part22_02}.
	
	\subsubsection{Covert Beamformer}
	By introducing an auxiliary variable ${\rm{I}_{\rm{R}}}$, the above problem can be reformulated in the following equivalent form:	
	\begin{subequations}\label{part22_03}
		\begin{align}
			\mathop {\max }\limits_{{{\bf{w}}_{\rm{R,0}}},{{\bf{w}}_{{\rm{R,1}}}},{\rm{I}_{\rm{R}}}}& {\rm{I}_{\rm{R}}}\\
			{\rm{s}}.{\rm{t}}. \quad	&  {\frac{{{{\left| \alpha  \right|^2}{\left| {{\bf{h}}_{\rm{T}}^H{{\bf{w}}_{{\rm{R,0}}}}} \right|}^2{{\left\| {{{\bf{h}}_{\rm{T}}}} \right\|}^2}}}}{{{\left| \alpha  \right|^2}{{\left| {{\bf{h}}_{\rm{T}}^H{{\bf{w}}_{{\rm{R,1}}}}} \right|}^2{{\left\| {{{\bf{h}}_{\rm{T}}}} \right\|}^2}} + \sigma _{\rm{R}}^2}}}   \ge {\rm{I}_{\rm{R}}},\\
			&{\left| {{\bf{h}}_{\rm{W}}^H{{\bf{w}}_{{\rm{R,1}}}}} \right|^2} = 0,\\
			&\frac{{{{\left| {{\bf{h}}_{\rm{B}}^H{{\bf{w}}_{{\rm{R,1}}}}} \right|}^2}}}{{{{\left| {{\bf{h}}_{\rm{B}}^H{{\bf{w}}_{{\rm{R,0}}}}} \right|}^2} + \sigma _{\rm{B}}^2}} \geq \beta ,\\
			&{\left\| {{{\bf{w}}_{{\rm{R}},{\rm{0}}}}} \right\|^2} + {\left\| {{{\bf{w}}_{{\rm{R}},{\rm{1}}}}} \right\|^2} \le {P_{{\rm{total}}}}.{\rm{ }}
		\end{align}
	\end{subequations}
	
 	Then, we use the semidefinite relaxation (SDR) approach \cite{Luo_SPM_2010} to relax problem $\eqref{part22_03}$ to
	\begin{subequations}\label{part22_13}
		\begin{align}
			&{{\bf{W}}_{\rm{R,0}}} = {{\bf{w}}_{\rm{R,0}}}{\bf{w}}_{\rm{R,0}}^H \Leftrightarrow {{\bf{W}}_{\rm{R,0}}}\underline  \succ {\bf{0}},{\rm{rank}}\left( {{{\bf{W}}_{\rm{R,0}}}} \right) = 1,\\
			&{{\bf{W}}_{{\rm{R,1}}}} = {{\bf{w}}_{{\rm{R,1}}}}{\bf{w}}_{{\rm{R,1}}}^H \Leftrightarrow {{\bf{W}}_{{\rm{R,1}}}}\underline  \succ
			{\bf{0}},{\rm{rank}}\left( {{{\bf{W}}_{{\rm{R,1}}}}} \right) = 1.
		\end{align}
	\end{subequations}
	
	By ignoring the rank 1
			constraints, we   get a relaxed version of problem $\eqref{part22_03}$ as follows
		\begin{align}
			\mathop {\max }\limits_{{{\bf{w}}_{\rm{R,0}}},{{\bf{w}}_{{\rm{R,1}}}},{\rm{I}_{\rm{R}}}}& {\rm{I}_{\rm{R}}}\label{part22_04}\\
			{\rm{s}}.{\rm{t}}. \quad	&{\left| \alpha  \right|^2}{\rm{Tr}}\left( {{\bf{h}}_{\rm{T}}^H{{\bf{W}}_{{\rm{R}},{\rm{0}}}}{{\bf{h}}_{\rm{T}}}} \right){\left\| {{{\bf{h}}_{\rm{T}}}} \right\|^2} \ge {{\rm{I}}_{\rm{R}}}\left( {{{\left| \alpha  \right|}^2}} \right. \nonumber \\
			&\qquad \left. { \times {\rm{Tr}}\left( {{\bf{h}}_{\rm{T}}^H{{\bf{W}}_{{\rm{R}},{\rm{1}}}}{{\bf{h}}_{\rm{T}}}} \right){{\left\| {{{\bf{h}}_{\rm{T}}}} \right\|}^2} + \sigma _{\rm{R}}^2} \right),\nonumber\\
			&{\rm{Tr}}\left( {{\bf{h}}_{\rm{B}}^H{{\bf{W}}_{{\rm{R,1}}}}{{\bf{h}}_{\rm{B}}}} \right) \ge \beta\left( {{\rm{Tr}}\left( {{\bf{h}}_{\rm{B}}^H{{\bf{W}}_{{\rm{R,0}}}}{{\bf{h}}_{\rm{B}}}} \right) + \sigma _{\rm{B}}^2} \right),\nonumber\\
			&{\rm{Tr}}\left( {{\bf{h}}_{\rm{W}}^H{{\bf{W}}_{{\rm{R,1}}}}{{\bf{h}}_{\rm{W}}}} \right) = 0,\nonumber\\
			&{\rm{Tr}}\left( {{{\bf{W}}_{{\rm{R}},{\rm{0}}}}} \right) + {\rm{Tr}}\left( {{{\bf{W}}_{\rm{R,1}}}} \right) \le {P_{{\rm{total}}}},\nonumber\\
			&{{\bf{W}}_{{\rm{R}},{\rm{0}}}}
			\underline  \succ{\bf{0}},{{\bf{W}}_{\rm{R,1}}}
			\underline  \succ{\bf{0}}.\nonumber
		\end{align}

	Note that, for any ${\rm{I}_{\rm{R}}} \ge 0$, it is a convex semidefinite program (SDP). Therefore, it is quasi-convex, and at any given ${\rm{I}_{\rm{R}}}$, by testing its feasibility, the optimal solution can be found.
	
	Therefore, we first covert problem $\eqref{part22_04}$ into a series of convex subproblems of ${\rm{I}_{\rm{R}}} \ge 0$, which can be solved by a standard convex optimization solver (such as CVX).
	Then, we use the bisection search method to find the proposed covert beamformers ${{\bf{W}}_{\rm{R,0}}}$ and ${{\bf{W}}_{\rm{R,1}}}$, which output the optimal solutions ${\bf{W}}_{\rm{R,0}}^*$ and ${\bf{W}}_{\rm{R,1}}^*$.
	
	\begin{algorithm}[htb]
		\caption{Bisection method for   problem \eqref{part22_04}}
		\label{alg:Bisection Method}
		\begin{algorithmic}[1]
			\State Determine the interval $\left[ {{\rm{ I}_{{\rm{R,l}}}},{\rm{ I}_{{\rm{R,end}}}}} \right]$, given the accuracy $\zeta_1   > 0$;
			\State Initialize ${\rm{ I}_{{\rm{R,l}}}}=0$, ${\rm{ I}_{{\rm{R,end}}}}={{\hat {\rm{I}}}_{\rm{R}}}$;
			
			\While {${{ \rm{ I}}_{{\rm{R,end}}}} - {{ \rm{ I}}_{{\rm{R,l}}}} \ge \zeta_1  $}
			\State  set ${{ \rm{ I}}_{{\rm{R}}}} = \left( {{{ \rm{ I}}_{{\rm{R,l}}}} + {{ \rm{ I}}_{{\rm{R,end}}}}} \right)/2$;
			\State {\bf{if}}    problem \eqref{part22_04}  is solvable, get  ${{\bf{W}}_{\rm{R,0}}}$ and ${{\bf{W}}_{{\rm{R}},1}}$,  then set ${{ \rm{ I}}_{{\rm{R,l}}}} = {{ \rm{ I}}_{{\rm{R}}}}$;
			\State {\bf{else}}, set ${{ \rm{ I}}_{{\rm{R,end}}}} = {{ \rm{ I}}_{{\rm{R,mid}}}}$;
			\EndWhile
			\State  Output   ${\bf{W}}_{{\rm{R}},1}^*$,${\bf{W}}_{\rm{R,0}}^*$;
		\end{algorithmic}
	\end{algorithm}
	
	Finally, we can  solve this problem based on the solutions given by Algorithm 1. {
		The   computational complexity of     Algorithm 1   is ${\cal O}\left( {\max {{\left\{ {4,2N} \right\}}^4}\sqrt {2N} \log \left( {{1 \mathord{\left/
						{\vphantom {1 {{\xi _1}}}} \right.
						\kern-\nulldelimiterspace} {{\zeta_1}}}} \right)\log \left( {{1 \mathord{\left/
						{\vphantom {1 \zeta_1  }} \right.
						\kern-\nulldelimiterspace} \zeta_1  }} \right)} \right)$, where    ${\zeta _1} > 0$ is the pre-defined accuracy of problem \eqref{part22_04} \cite{Luo_SPM_2010,Grant09cvx,Sturm06SeDuMi}.}
	However,  because of the rank relaxation of SDR,
	the ranks of
	the optimal solutions   ${\bf{W}}_{\rm{R,0}}^*$ and ${\bf{W}}_{\rm{R,1}}^*$   may larger than 1.
	When ${\rm{rank}}\left( {{\bf{W}}_{\rm{R,0}}^*} \right) = 1$  and ${\rm{rank}}\left( {{\bf{W}}_{\rm{R,1}}^*} \right) = 1$, we employ the singular value decomposition   to decompose ${\bf{W}}_{\rm{R,0}}^*$ and ${\bf{W}}_{\rm{R,1}}^*$, i.e., ${{\bf{W}}_{\rm{R,0}}} = {{\bf{w}}_{\rm{R,0}}}{\bf{w}}_{\rm{R,0}}^H$ and ${{\bf{W}}_{{\rm{R,1}}}} = {{\bf{w}}_{{\rm{R,1}}}}{\bf{w}}_{{\rm{R,1}}}^H$. On the other hand, when ${\rm{rank}}\left( {{\bf{W}}_{\rm{R,0}}^*} \right) > 1$ or ${\rm{rank}}\left( {{\bf{W}}_{\rm{R,1}}^*} \right) > 1$, we apply   the Gaussian randomization procedure \cite{Luo_SPM_2010} to problem $\eqref{part22_03}$ and  get  high-quality rank $1$ beamformers.


	\subsubsection{Zero-Forcing Beamformer}
To eliminate interference signals of Willie and the radar,
	 we design zero-forcing beamformer ${{\bf{w}}_{\rm{R,1}}}$ satisfying ${\bf{h}}_{{\rm{W}}}^H{{\bf{w}}_{\rm{R,1}}} = 0$ and ${\bf{h}}_{\rm{T}}{\bf{h}}_{{\rm{T}}}^H{{\bf{w}}_{\rm{R,1}}} = \bf{0}$.
	In addition, we  design ${{\bf{w}}_{{\rm{R,0}}}}$ to eliminate Bob's interference, i.e., ${\bf{h}}_{{\rm{B}}}^H{{\bf{w}}_{{\rm{R,0}}}} = 0$. Then, problem $\eqref{part22_02}$ can be expressed as
	\begin{subequations}\label{part22_11}
		\begin{align}
			\mathop {\max }\limits_{{{\bf{w}}_{{\rm{R}},{\rm{0}}}},{{\bf{w}}_{{\rm{R}},{\rm{1}}}}} &{\left| {{\bf{h}}_{\rm{T}}^H{{\bf{w}}_{{\rm{R}},0}}} \right|^2}\label{part22_11_1}\\
			{\rm{s}}{\rm{.t}}{\rm{.}} \quad &{\bf{h}}_{\rm{T}}^H{{\bf{w}}_{{\rm{R}},{\rm{1}}}} = 0,\label{part22_11_2}\\
			&{\bf{h}}_{\rm{W}}^H{{\bf{w}}_{{\rm{R}},{\rm{1}}}} = 0,\label{part22_11_3}\\
			&{\bf{h}}_{{\rm{B}}}^H{{\bf{w}}_{{\rm{R,0}}}} = 0,\label{part22_11_4}\\
			&{\left| {{\bf{h}}_{\rm{B}}^H{{\bf{w}}_{{\rm{R}},{\rm{1}}}}} \right|^2} \ge \beta \sigma _{\rm{B}}^{\rm{2}} ,\label{part22_11_7} \\
			&{\left\| {{{\bf{w}}_{{\rm{R}},{\rm{0}}}}} \right\|^2} + {\left\| {{{\bf{w}}_{\rm{R,1}}}} \right\|^2} \le {P_{{\rm{total}}}}.\label{part22_11_6}
		\end{align}
	\end{subequations}
	
	To solve   problem \eqref{part22_11}, we first optimize the beamformer ${{{\bf{w}}_{{\rm{R}},{\rm{1}}}}}$ by minimizing the transmission power ${\left\| {{{\bf{w}}_{{\rm{R}},{\rm{1}}}}} \right\|^2}$ under the constraints   $\eqref{part22_11_2}$, $\eqref{part22_11_3}$ and $\eqref{part22_11_7}$.
	This is because the value of the objective function $\eqref{part22_11_1}$       increases as the power of  ${{\bf{w}}_{\rm{R,0}}}$ increases, but
 does not depend on ${{{\bf{w}}_{{\rm{R}},{\rm{1}}}}}$.
	Therefore, in order to maximize the objective function $\eqref{part22_11_1}$, it is necessary to design the beamformer ${{\bf{w}}_{\rm{R,1}}}$ with the least transmission power.
	Therefore, the design problem of the ZF beamformer ${{\bf{w}}_{\rm{R,1}}}$ can be expressed as
	\begin{align}\label{part22_12}
		\mathop {\min }\limits_{{{\bf{w}}_{{\rm{R,1}}}}} &{\left\| {{{\bf{w}}_{{\rm{R}},{\rm{1}}}}} \right\|^2}\\
		& \eqref{part22_11_2}, \eqref{part22_11_3}, \eqref{part22_11_7},\notag
	\end{align}
	which is a non-convex problem.
	
	To solve this problem, we adopt the SDR approach to relax problem $\eqref{part22_12}$. Therefore, using $\eqref{part22_13}$ and ignoring the rank 1 constraint, problem $\eqref{part22_12}$ can be reformulated as
	\begin{subequations}\label{part22_14}
		\begin{align}
			\mathop {\min }\limits_{{{\bf{w}}_{{\rm{R,1}}}}} &{\rm{Tr}}\left( {{{\bf{W}}_{{\rm{R}},{\rm{1}}}}} \right)\\
			{\rm{s}}{\rm{.t}}{\rm{.}}
			&{\rm{Tr}}\left( {{\bf{h}}_{{\rm{T}}}^H}{{\bf{W}}_{{\rm{R}},{\rm{1}}}}{{\bf{h}}_{{\rm{T}}}} \right) = 0,\\
			&{\rm{Tr}}\left( {{\bf{h}}_{{\rm{W}}}^H}{{\bf{W}}_{{\rm{R}},{\rm{1}}}}{{\bf{h}}_{{\rm{W}}}} \right) = 0,\\
			&{\rm{Tr}}\left( {{\bf{h}}_{\rm{B}}^H{{\bf{W}}_{{\rm{R,1}}}}{{\bf{h}}_{\rm{B}}}} \right) \geq \beta \sigma _{\rm{B}}^2,\\
			&{{\bf{W}}_{{\rm{R}},{\rm{0}}}}\succeq{\bf{1}}.
		\end{align}
	\end{subequations}

	However, due to relaxed conditions, the rank of ${\bf{W}}_{{\rm{R,1}}}^{{\rm{opt}}}$ may not be equal to 1, where   ${\bf{W}}_{{\rm{R,1}}}^{{\rm{opt}}}$   denotes the optimal solution of problem $\eqref{part22_14}$.
	If ${\rm{rank}}\left( {{\bf{W}}_{{\rm{R,1}}}^{{\rm{opt}}}} \right) = 1$,  ${\bf{W}}_{{\rm{R,1}}}^{{\rm{opt}}}$ is the optimal solution, and the optimal beamformer ${{\bf{w}}_{{\rm{R}},{\rm{1}}}}$ can be obtained using SVD, i.e., ${\bf{W}}_{{\rm{R, 1}}}^{{\rm{opt}}} = {{\bf{w}}_{{\rm{R}},{\rm{1}}}}{\bf{w}}_ {{\rm{R,1}}}^H$. Otherwise, if ${\rm{rank}}\left( {{\bf{W}}_{{\rm{R,1}}}^ {{\rm{opt}}}} \right)> 1$, the Gaussian randomization process can be used to obtain the high-quality rank 1 solution of problem $\eqref{part22_14}$.
	 Therefore, problem $ \eqref{part22_11}$ can be expressed as
	\begin{subequations}\label{part22_15}
		\begin{align}
			\mathop {\max }\limits_{{{\bf{w}}_{\rm{R,0}}}} &{\left| {{\bf{h}}_{{\rm{T}}}^H{{\bf{w}}_{\rm{R,0}}}} \right|^2}\\
			{\rm{s}}{\rm{.t}}{\rm{.}}  &{\left\| {{{\bf{w}}_{\rm{R,0}}}} \right\|^2} + {P_{\rm{R}}} \le {P_{{\rm{total}}}},\\
			&{\bf{h}}_{{\rm{B}}}^H{{\bf{w}}_{\rm{R,0}}} = 0,
		\end{align}
	\end{subequations}
	where 	  ${P_{\rm{R}}} = {\left\| {{\bf{W}}_{{\rm{R,1}}}^{{\rm{opt}}} } \right\|^2}$   denotes the transmission power of ${\bf{W}}_{{\rm{R,1}}}^{{\rm{opt}}}$.
	After   simplifications, we obtain
	\begin{subequations}\label{part22_16}
		\begin{align}
			\mathop {\max }\limits_{{{\bf{w}}_{\rm{R,0}}}}& {\mathop{\rm Re}\nolimits} \left\{ {{\bf{h}}_{{\rm{T}}}^H{{\bf{w}}_{\rm{R,0}}}} \right\}\\
			{\rm{s}}{\rm{.t.}}&{\rm{Im}}\left\{ {{\bf{h}}_{{\rm{T}}}^H{{\bf{w}}_{\rm{R,0}}}} \right\} = 0,\\
			&{\left\| {{{\bf{w}}_{\rm{R,0}}}} \right\|^2} + {P_{\rm{R}}} \le {P_{{\rm{total}}}},\\
			&{\bf{h}}_{{\rm{B}}}^H{{\bf{w}}_{\rm{R,0}}} = 0.
		\end{align}
	\end{subequations}
	
	Then, problem $\eqref{part22_16}$ is a SOCP, which can be optimized with standard convex optimization solvers (such as CVX).
	\subsection{Rate Maximization}
	In the perfect WCSI scenario, problem (20) can be mathematically formulated as
	\begin{subequations}\label{part2_01}
		\begin{align}
			\mathop {\max }\limits_{{{\bf{w}}_{{\rm{R,0}}}},{{\bf{w}}_{{\rm{R,1}}}}}& {R_{\rm{B}}}\left( {{{\bf{w}}_{{\rm{R}},{\rm{0}}}},{{\bf{w}}_{{\rm{R}},{\rm{1}}}}} \right)\\
			{\rm{s}}.{\rm{t}}. \quad &D\left( {{p_0}\left\| {{p_1}} \right.} \right) = 0,\label{part2_013}\\
			&{\rm{I}}\left( {{\bf{y}_{\rm{R}}};{{\bf{h}}_{\rm{T}}}\left| {{s_{\rm{R}}}} \right.} \right) \geq \gamma ,\\
			&{\left\| {{{\bf{w}}_{{\rm{R}},{\rm{0}}}}} \right\|^2} + {\left\| {{{\bf{w}}_{{\rm{R}},{\rm{1}}}}} \right\|^2} \le {P_{{\rm{total}}}}.
		\end{align}
	\end{subequations}
	
	Combining with $\eqref{part1_05}$, we obtain the following equivalent from

		\begin{align}
			\mathop {\max }\limits_{{{\bf{w}}_{{\rm{R,0}}}},{{\bf{w}}_{{\rm{R,1}}}}}& \frac{{{{\left| {{\bf{h}}_{\rm{B}}^H{{\bf{w}}_{{\rm{R,1}}}}} \right|}^2}}}{{{{\left| {{\bf{h}}_{\rm{B}}^H{{\bf{w}}_{{\rm{R,0}}}}} \right|}^2} + \sigma _{\rm{B}}^2}}\label{part2_02}\\
			{\rm{s}}.{\rm{t}}. \quad	&
			{\left| {{\bf{h}}_{\rm{W}}^H{{\bf{w}}_{{\rm{R,1}}}}} \right|^2} = 0,\nonumber\\
			&  \frac{1}{2}\log \left( {1 + \frac{{{\left| \alpha  \right|^2}{{\left| {{\bf{h}}_{\rm{T}}^H{{\bf{w}}_{{\rm{R,0}}}}} \right|}^2}{{\left\| {{{\bf{h}}_{\rm{T}}}} \right\|}^2}}}{{{\left| \alpha  \right|^2}{{\left| {{\bf{h}}_{\rm{T}}^H{{\bf{w}}_{{\rm{R,1}}}}} \right|}^2}{{\left\| {{{\bf{h}}_{\rm{T}}}} \right\|}^2} + \sigma _{\rm{R}}^2}}} \right) \geq \gamma ,\nonumber\\
			&{\left\| {{{\bf{w}}_{{\rm{R}},{\rm{0}}}}} \right\|^2} + {\left\| {{{\bf{w}}_{{\rm{R}},{\rm{1}}}}} \right\|^2} \le {P_{{\rm{total}}}}.\nonumber
		\end{align}

	Similar to the previous subsection, we apply the binary search
	method and ZF beamforming design to solve problem $\eqref{part2_02}$. For the sake of brevity,
	we omit the derivation details.
	\section{Beamforming Design based on Imperfect WCSI}
	Generally speaking, the WCSI may not be always accessible for the radar because of the potential limited cooperation between the radar and Willie. {{Therefore, we consider a more practical application scenario, in which Willie is an ordinary user and the radar does not know the full WCSI ${{\bf{h}}_{{\rm{W}}}}$ {{\cite{Shahzad_VTC_2017,Vakili_ICASSPP_2006,Li_CORR_2020,Li_ITC_2021}}}}}. In this case, the imperfect $	 {{\bf{h}}_{{\rm{W}}}}$ is modeled as
	\begin{subequations}
		\begin{align}
			{{\bf{h}}_{{\rm{W}}}} = {\widehat {\bf{h}}_{{\rm{W}}}} + \Delta {{\bf{h}}_{{\rm{W}}}},
		\end{align}
	\end{subequations}
	where ${\widehat {\bf{h}}_{{\rm{W}}}}$ denotes the estimated CSI, and $\Delta {{\bf{h}}_{{\rm{W}}}}$ denotes corresponding CSI error vector. Moreover, the CSI error $\Delta {{\bf{h}}_{{\rm{W}}}}$ is characterized by an ellipsoidal region, i.e.,
	\begin{subequations}
		\begin{align}
			{\mathcal{E} _{{\rm{W}}}} \buildrel \Delta \over = \left\{ {\Delta {{\bf{h}}_{{\rm{W}}}}\left| {\Delta {\bf{h}}_{{\rm{W}}}^H{{\bf{C}}_{{\rm{W}}}}\Delta {{\bf{h}}_{{\rm{W}}}} \le {\upsilon _{{\rm{W}}}}} \right.} \right\},
		\end{align}
	\end{subequations}
	where $\begin{array}{l}	
		{{\bf{C}}_{{\rm{W}}}} = {\bf{C}}_{{\rm{W}}}^H\succeq{\bf{0}}
	\end{array}$ controls the axes of the ellipsoid, and ${\upsilon _{{\rm{W}}}} > 0$ which determines the volume of the ellipsoid \cite{Forouzesh_TVT_2020,He_TIFS_2013}.
	Unfortunately, since the radar does not know the full WCSI, perfect covert transmission $\eqref{D}$  is difficult to achieve. Therefore, we
	adopt $\eqref{part1_07}$ as covertness
	constraints \cite{Yan_TIFS_2019,Yan_TWC_2019}.
	\subsection{MI Maximization}
	\subsubsection{Case of $D\left( {{p_0}\left\| {{p_1}} \right.} \right) \le 2{\varepsilon ^2}$}
	The optimization problem $\eqref{C2}$ can be written as
	\begin{subequations}\label{part32_01}
		\begin{align}
			\mathop {\max }\limits_{{{\bf{w}}_{{\rm{R,0}}}},{{\bf{w}}_{{\rm{R,1}}}}} &  {\frac{{{{\left| \alpha  \right|^2}{\left| {{\bf{h}}_{\rm{T}}^H{{\bf{w}}_{{\rm{R,0}}}}} \right|}^2{{\left\| {{{\bf{h}}_{\rm{T}}}} \right\|}^2}}}}{{{\left| \alpha  \right|^2}{{\left| {{\bf{h}}_{\rm{T}}^H{{\bf{w}}_{{\rm{R,1}}}}} \right|}^2{{\left\| {{{\bf{h}}_{\rm{T}}}} \right\|}^2}} + \sigma _{\rm{R}}^2}}}   \\
			{\rm{s}}.{\rm{t}}.\quad &D\left( {{p_0}\left\| {{p_1}} \right.} \right) \le 2{\varepsilon ^2},\label{part32_01_1}\\
			&\frac{{{{\left| {{\bf{h}}_{\rm{B}}^H{{\bf{w}}_{{\rm{R,1}}}}} \right|}^2}}}{{{{\left| {{\bf{h}}_{\rm{B}}^H{{\bf{w}}_{{\rm{R,0}}}}} \right|}^2} + \sigma _{\rm{B}}^2}} \geq \beta ,\\
			&{\left\| {{{\bf{w}}_{{\rm{R}},{\rm{0}}}}} \right\|^2} + {\left\| {{{\bf{w}}_{{\rm{R}},{\rm{1}}}}} \right\|^2} \le {P_{{\rm{total}}}},\\
			&{{\bf{h}}_{{\rm{W}}}} = {\widehat {\bf{h}}_{{\rm{W}}}} + \Delta {{\bf{h}}_{{\rm{W}}}}.\label{part32_01_2}
		\end{align}
	\end{subequations}
	
	It is clear that the problem  \eqref{part32_01}  is  non-convex, and therefore it is difficult to obtain the optimal solution directly. To deal with this issue, we first reformulate the covertness constraint $\eqref{part32_01_1}$ by exploiting the property of function $f\left( x \right) = \ln x + \frac{1}{x} - 1$ for $x > 0$. Specifically, the covertness constant $D\left( {{p_0}\left\| {{p_1}} \right.} \right) = \ln \frac{{{\lambda _1}}}{{{\lambda _0}}} + \frac{{{\lambda _0}}}{{{\lambda _1}}} - 1 \le 2{\varepsilon ^2}$ can be equivalently transformed as	
	\begin{align}
		\bar a \le \frac{{{\lambda _1}}}{{{\lambda _0}}} \le \bar b{\kern 1pt} {\kern 1pt} ,
	\end{align}
	where $\bar a$ and $\bar b$ are the two roots of the equation $\ln \frac{{{\lambda _1}}}{{{\lambda _0}}} + \frac{{{\lambda _0}}}{{{\lambda _1}}} - 1 = 2{\varepsilon ^2}$. Therefore, we may reformulate $\eqref{part32_01_1}$ as
	\begin{align}\label{part3_02}
		\bar a \le \frac{{{{\left| {{\bf{h}}_{\rm{W}}^H{{\bf{w}}_{{\rm{R}},{\rm{0}}}}} \right|}^2} + {{\left| {{\bf{h}}_{\rm{W}}^H{{\bf{w}}_{{\rm{R,1}}}}} \right|}^2} + \sigma _{\rm{W}}^2}}{{{{\left| {{\bf{h}}_{\rm{W}}^H{{\bf{w}}_{{\rm{R}},{\rm{0}}}}} \right|}^2} + \sigma _{\rm{W}}^2}} \le \bar b.
	\end{align}
	
	For constraint $\eqref{part32_01_2}$, since $\Delta {{\bf{h}}_{{\rm{W}}}} \in {{\cal E}_{{\rm{W}}}}$, there are infinite choices for $\Delta {{\bf{h}}_{{\rm{W}}}}$, which makes the problem $\eqref{part32_01}$ non-convex and intractable. To overcome this challenge, we define ${{\bf{W}}_{\rm{R,0}}}  =  {{\bf{w}}_{\rm{R,0}}}{\bf{w}}_{{\rm{R}},0}^H$,    ${{\bf{W}}_{{\rm{R}},1}}  =  {{\bf{w}}_{{\rm{R}},1}}{\bf{w}}_{{\rm{R}},1}^H$, $\widehat {\bf{W}}_1 \buildrel \Delta \over = \left( {1 - \bar a} \right){{\bf{W}}_{{\rm{R,0}}}} + {{\bf{W}}_{{\rm{R,1}}}}$ and $\widehat {\bf{W}}_2 \buildrel \Delta \over = \left( {1 - \bar b} \right){{\bf{W}}_{{\rm{R,0}}}} + {{\bf{W}}_{{\rm{R,1}}}}$.
	Then,  constraint \eqref{part32_02} can be equivalently reexpressed as
	\begin{subequations}
		\begin{align}
			&\Delta {\bf{h}}_{\rm{W}}^H\widehat {\bf{W}}_1\Delta {{\bf{h}}_{\rm{W}}} + 2\Delta {\bf{h}}_{\rm{W}}^H\widehat {\bf{W}}_1{{{\bf{\hat h}}}_{\rm{W}}} + {\bf{\hat h}}_{\rm{W}}^H\widehat {\bf{W}}_1{{{\bf{\hat h}}}_{\rm{W}}} \ge \sigma _{\rm{W}}^2\left( {\bar a - 1} \right)\label{W1a},\\
			& \Delta {\bf{h}}_{\rm{W}}^H\widehat {\bf{W}}_2\Delta {{\bf{h}}_{\rm{W}}} + 2\Delta {\bf{h}}_{\rm{W}}^H\widehat {\bf{W}}_2{{{\bf{\hat h}}}_{\rm{W}}} + {\bf{\hat h}}_{\rm{W}}^H\widehat {\bf{W}}_2{{{\bf{\hat h}}}_{\rm{W}}}\le \sigma _{\rm{w}}^2\left( {\bar b - 1} \right)\label{W1b},
		\end{align}
	\end{subequations}
	
	Furthermore, by ignoring the rank-one constraints of ${{\bf{W}}_{\rm{R,0}}}$ and ${{\bf{W}}_{\rm{R,1}}}$, the SDR of problem$\eqref{part32_01}$ can be reformulated as
	\begin{subequations}\label{part3_03}
		\begin{align}
			\mathop {\max }\limits_{{{\bf{w}}_{{\rm{R,0}}}},{{\bf{w}}_{{\rm{R,1}}}},{\rm{{\tilde I}}_{\rm{R}}}}& {\rm{{\tilde I}}_{\rm{R}}}\\
			{\rm{s}}.{\rm{t}}. \quad&{\left| \alpha  \right|^2}{\rm{Tr}}\left( {{\bf{h}}_{\rm{T}}^H{{\bf{W}}_{{\rm{R}},{\rm{0}}}}{{\bf{h}}_{\rm{T}}}} \right){\left\| {{{\bf{h}}_{\rm{T}}}} \right\|^2} \ge {{{\rm{\tilde I}}}_{\rm{R}}}\left( {{{\left| \alpha  \right|}^2}} \right. \nonumber\\
			&\qquad \left. { \times {\rm{Tr}}\left( {{\bf{h}}_{\rm{T}}^H{{\bf{W}}_{{\rm{R}},{\rm{1}}}}{{\bf{h}}_{\rm{T}}}} \right){{\left\| {{{\bf{h}}_{\rm{T}}}} \right\|}^2} + \sigma _{\rm{R}}^2} \right), \label{equal2}\\
			&{\rm{Tr}}\left( {{\bf{h}}_{\rm{B}}^H{{\bf{W}}_{{\rm{R,1}}}}{{\bf{h}}_{\rm{B}}}} \right) \ge \beta \left( {{\rm{Tr}}\left( {{\bf{h}}_{\rm{B}}^H{{\bf{W}}_{{\rm{R,0}}}}{{\bf{h}}_{\rm{B}}}} \right) + \sigma _{\rm{B}}^2} \right),\label{equal1}\\
			& {\Delta {\bf{h}}_{{\rm{W}}}^H{{\bf{C}}_{{\rm{W}}}}\Delta {{\bf{h}}_{{\rm{W}}}} \le {\upsilon _{{\rm{W}}}}},\\
			&{\rm{Tr}}\left( {{{\bf{W}}_{{\rm{R}},{\rm{0}}}}} \right) + {\rm{Tr}}\left( {{{\bf{W}}_{{\rm{R,1}}}}} \right) \le {P_{{\rm{total}}}},\label{equal3}\\
			 &{{\bf{W}}_{\rm{R,0}}}\succeq{\bf{0}},{{\bf{W}}_{\rm{R,1}}}\succeq{\bf{0}},\label{equal4}\\
			&	\eqref{W1a},\eqref{W1b}.\notag
		\end{align}
	\end{subequations}
	
	Here, $\Delta {{\bf{h}}_{{\rm{W}}}} \in {{\cal E}_{{\rm{W}}}}$ involves an infinite number of constraints, which makes problem $\eqref{part3_03}$ still computationally prohibitive. We apply the S-lemma to transform the constraints into a certain set of linear matrix inequalities (LMIs), which is a  tractable safe approximation.
	
	\textbf{Lemma 1 (S-Procedure \cite{Ng_TWC_2014})}: Let a function ${f_m}\left( x \right),m \in \left\{ {1,2} \right\},x \in {{\mathbb{C}}^{N \times 1}}$, be defined as
	\begin{align}
		{f_m}\left( x \right) = {{\bf{x}}^H}{{\bf{A}}_m}{\bf{x}} + 2\text{Re}\left\{ {{\bf{b}}_m^{\rm{H}}{\bf{x}}} \right\} + {c_m},
	\end{align}
	where ${{\bf{A}}_m} \in {{\mathbb{C}}^N}$ is a complex
	Hermitian matrix, ${{\bf{b}}_m} \in {{\mathbb{C}}^{N \times 1}}$ and ${c_m} \in {{\mathbb{R}}^{1 \times 1}}$. Then, the
	implication relation ${f_1}\left( x \right) \le 0 \Rightarrow {f_2}\left( x \right) \le 0$ holds if and  {only if there
		exists a variable $\eta   \ge 0$ } such that
	\begin{align}
		{\eta } \left[ {\begin{array}{*{20}{c}}
				{{{\bf{A}}_1}}&{{{\bf{b}}_1}}\\
				{{\bf{b}}_1^{{H}}}&{{c_1}}
		\end{array}} \right] - \left[ {\begin{array}{*{20}{c}}
				{{{\bf{A}}_2}}&{{{\bf{b}}_2}}\\
				{{\bf{b}}_2^{{H}}}&{{c_2}}
		\end{array}} \right]\underline  \succ  {\bf{0}}\label{spro}.
	\end{align}
	
	Consequently, by applying Lemma 1, constraints $\eqref{W1a}$ and $\eqref{W1b}$ can be, respectively, reformulated as the following finite number of LMIs:	
	\begin{subequations}
		\begin{align}
			&\left[ {\begin{array}{*{20}{c}}
					{{{\widehat {\bf{W}}}_1} + {\eta _1}{{\bf{C}}_{\rm{W}}}}&{{{\widehat {\bf{W}}}_1}{{{\bf{\hat h}}}_{\rm{W}}}}\\
					{{\bf{\hat h}}_{\rm{W}}^H{{\widehat {\bf{W}}}_1}}&{{\bf{\hat h}}_{\rm{W}}^H{{\widehat {\bf{W}}}_1}{{{\bf{\hat h}}}_{\rm{W}}} - \sigma _{\rm{W}}^2\left( {\bar a - 1} \right) - {\eta _1}{v_{\rm{W}}}}
			\end{array}} \right]\underline  \succ  {\bf{0}}\label{slemma1},\\
			&\left[ {\begin{array}{*{20}{c}}
					{ - \widehat {\bf{W}}_2 + {\eta _2}{{\bf{C}}_{\rm{W}}}}&{ - \widehat {\bf{W}}_2{{{\bf{\hat h}}}_{\rm{W}}}}\\
					{ - {\bf{\hat h}}_{\rm{W}}^H\widehat {\bf{W}}_2}&{ - {\bf{\hat h}}_{\rm{W}}^H\widehat {\bf{W}}_2{{{\bf{\hat h}}}_{\rm{W}}} + \sigma _{\rm{W}}^2\left( {\bar b - 1} \right)   - {\eta _2}{v_{\rm{W}}}}
			\end{array}} \right]\underline  \succ  {\bf{0}}\label{slemma2}.
		\end{align}
	\end{subequations}
	
	Thus, we obtain a conservative approximation of problem $\eqref{part3_03}$ as follows:
	 		\begin{align}
			\mathop {\max }\limits_{{{\bf{w}}_{{\rm{R,0}}}},{{\bf{w}}_{{\rm{R,1}}}},{{\rm{{\tilde I}}}_{\rm{R}}}}& {{\rm{{\tilde I}}}_{\rm{R}}}\label{part3_04}\\
			{\rm{s}}.{\rm{t}}.\quad & \eqref{equal2},\eqref{equal1},\eqref{slemma1},\eqref{slemma2},\eqref{equal3},\eqref{equal4}.\notag
		\end{align}

	When ${{\rm{{\tilde I}}}_{\rm{R}}}$ is fixed, problem $\eqref{part3_04}$ is a convex SDP, which can be efficiently solved by off-the-shelf convex solvers.	Therefore, problem $\eqref{part3_04}$ can be efficiently solved by the
	proposed bisection method, which is summarized in Algorithm
	2. {
		The   computational complexity of     Algorithm 2   is ${\cal O}\left( {\max {{\left\{ {5,2N - 1} \right\}}^4}\sqrt {2N - 1} \log \left( {{1 \mathord{\left/
						{\vphantom {1 {{\xi _2}}}} \right.
						\kern-\nulldelimiterspace} {{\zeta_2}}}} \right)\log \left( {{1 \mathord{\left/
						{\vphantom {1 \zeta_2  }} \right.
						\kern-\nulldelimiterspace} {\zeta_2} }} \right)} \right)$, where    ${\zeta_2} > 0$ is the pre-defined accuracy of problem \eqref{part3_04}.}
	
	\begin{algorithm}[htb]
		\caption{Bisection method for   problem  \eqref{part3_04}}
		\label{alg:2}
		\begin{algorithmic}[1]
			\State Determine the interval $\left[ {{{\rm{{\tilde I}}}_{{\rm{R,l}}}},{{\rm{{\tilde I}}}_{{\rm{R,end}}}}} \right]$, given the accuracy $\zeta_2  > 0$;
			\State Initialize ${{\rm{{\tilde I}}}_{{\rm{R,l}}}}=0$, ${{\rm{{\tilde I}}}_{{\rm{R,end}}}}={\rm{{\hat I}}_{\rm{R}}}$;
			\While {${{\rm{{\tilde I}}}_{{\rm{R,end}}}} - {{\rm{{\tilde I}}}_{{\rm{R,l}}}} \ge \zeta_2$}
			\State  Let ${{\rm{{\tilde I}}}_{{\rm{R,mid}}}} = \left( {{{\rm{{\tilde I}}}_{{\rm{R,l}}}} + {{\rm{{\tilde I}}}_{{\rm{R,end}}}}} \right)/2$;
			\State  {\bf{if}}    problem \eqref{part3_04} is solvable, we obtain  ${{\bf{W}}_{\rm{R,0}}}$ and ${{\bf{W}}_{{\rm{R}},1}}$, and set ${{\rm{{\tilde I}}}_{{\rm{R,l}}}} = {{\rm{{\tilde I}}}_{{\rm{R,mid}}}}$;
			\State {\bf{else}}, let ${{\rm{{\tilde I}}}_{{\rm{R,end}}}} = {{\rm{{\tilde I}}}_{{\rm{R,mid}}}}$;
			\EndWhile
			\State Output the optimal solutions ${\bf{W}}_{{\rm{R}},0}^*$,${\bf{W}}_{\rm{R,1}}^*$.
		\end{algorithmic}
	\end{algorithm}
	\subsubsection{Case of $D\left( {{p_1}\left\| {{p_0}} \right.} \right) \le 2{\varepsilon ^2}$}
	In this subsection, we consider the constraint $D\left( {{p_1}\left\| {{p_0}} \right.} \right) \le 2{\varepsilon ^2}$, where the   corresponding robust covert rate maximization problem  can be   formulated as
	\begin{subequations}\label{part32_02}
		\begin{align}
			\mathop {\max }\limits_{{{\bf{w}}_{{\rm{R,0}}}},{{\bf{w}}_{{\rm{R,1}}}}} &  {\frac{{{{\left| \alpha  \right|^2}{\left| {{\bf{h}}_{\rm{T}}^H{{\bf{w}}_{{\rm{R,0}}}}} \right|}^2{{\left\| {{{\bf{h}}_{\rm{T}}}} \right\|}^2}}}}{{{{\left| \alpha  \right|^2}{\left| {{\bf{h}}_{\rm{T}}^H{{\bf{w}}_{{\rm{R,1}}}}} \right|}^2{{\left\| {{{\bf{h}}_{\rm{T}}}} \right\|}^2}} + \sigma _{\rm{R}}^2}}}   \\
			{\rm{s}}.{\rm{t}}.\quad &D\left( {{p_1}\left\| {{p_0}} \right.} \right) \le 2{\varepsilon ^2},\label{part32_02_1}\\
			&\frac{{{{\left| {{\bf{h}}_{\rm{B}}^H{{\bf{w}}_{{\rm{R,1}}}}} \right|}^2}}}{{{{\left| {{\bf{h}}_{\rm{B}}^H{{\bf{w}}_{{\rm{R,0}}}}} \right|}^2} + \sigma _{\rm{B}}^2}} \geq \beta ,\\
			&{\left\| {{{\bf{w}}_{{\rm{R}},{\rm{0}}}}} \right\|^2} + {\left\| {{{\bf{w}}_{{\rm{R}},{\rm{1}}}}} \right\|^2} \le {P_{{\rm{total}}}},\\
			&{{\bf{h}}_{{\rm{W}}}} = {\widehat {\bf{h}}_{{\rm{W}}}} + \Delta {{\bf{h}}_{{\rm{W}}}}{{,}}\label{part32_02_2}
		\end{align}
	\end{subequations}
	where $D\left( {{p_1}\left\| {{p_0}} \right.} \right) = \ln \frac{{{\lambda _0}}}{{{\lambda _1}}} + \frac{{{\lambda _1}}}{{{\lambda _0}}} - 1$.
	
	Note that, problem \eqref{part32_02}
	is similar to problem \eqref{part32_01} except for the covertness constraint.
	The covertness constraint $D\left( {{p_1}\left\| {{p_0}} \right.} \right) = \ln \frac{{{\lambda _0}}}{{{\lambda _1}}} + \frac{{{\lambda _1}}}{{{\lambda _0}}} - 1 \le 2{\varepsilon ^2}$ can be  equivalently transformed as
	\begin{align}
		\bar c \le \frac{{{\lambda _0}}}{{{\lambda _1}}} \le \bar d,
	\end{align}
	where $\bar c= {\bar a}$ and $\bar d= {\bar b}$, are the two roots of the equation $\ln \frac{{{\lambda _0}}}{{{\lambda _1}}} + \frac{{{\lambda _1}}}{{{\lambda _0}}} - 1 = 2{\varepsilon ^2}$.
	Similar to the previous subsection, we apply the relaxation and restriction approach to solve   problem $ \eqref{part32_02}$. For the sake of brevity, we omit the detailed derivations. Although the methods used in the two scenarios are the same, the achievable covert rates are quite different under the two different signal constraints. We will illustrate and discuss this issue in the next section.
	\subsection{Rate Maximization}
	
	In practical applications, the obtained WCSI is often degraded by  estimation errors. Therefore, we further investigate  robust beamforming design for  problem $\eqref{part2_01}$. In this case, it is difficult to achieve perfect covert transmission, namely $D\left( {{p_0}\left\| {{p_1}} \right.} \right) = 0$. Hence, we use $D\left( {{p_0}\left\| {{p_1}} \right.} \right) \le 2{\varepsilon ^2}$ and $D\left( {{p_1}\left\| {{p_0}} \right.} \right) \le 2{\varepsilon ^2}$ given by $\eqref{part1_07}$ as hidden constraints.
		Similar to the previous subsection, we can apply the relaxation and restriction approach, and the detailed derivations are omitted for brevity. The design results will be numerically discussed in the next section.

	\section{Numerical Results}
	In this section, we present the  numerical results   to demonstrate the effectiveness of the proposed beamforming
	design schemes. Without loss of
	generality, we assume that  $\sigma _{\rm{B}}^2=\sigma _{\rm{W}}^2=\sigma _{\rm{T}}^2=0 \rm{dBm}$ \cite{Liu_RC_2020}, the path-loss coefficient $\alpha = 1$ \cite{Cheng_TSP_2019}.
	\subsection{Perfect WCSI}
	
	\begin{figure}[htbp]
		\begin{minipage}[t]{0.45\textwidth}
			\centering
			\includegraphics[width=7cm]{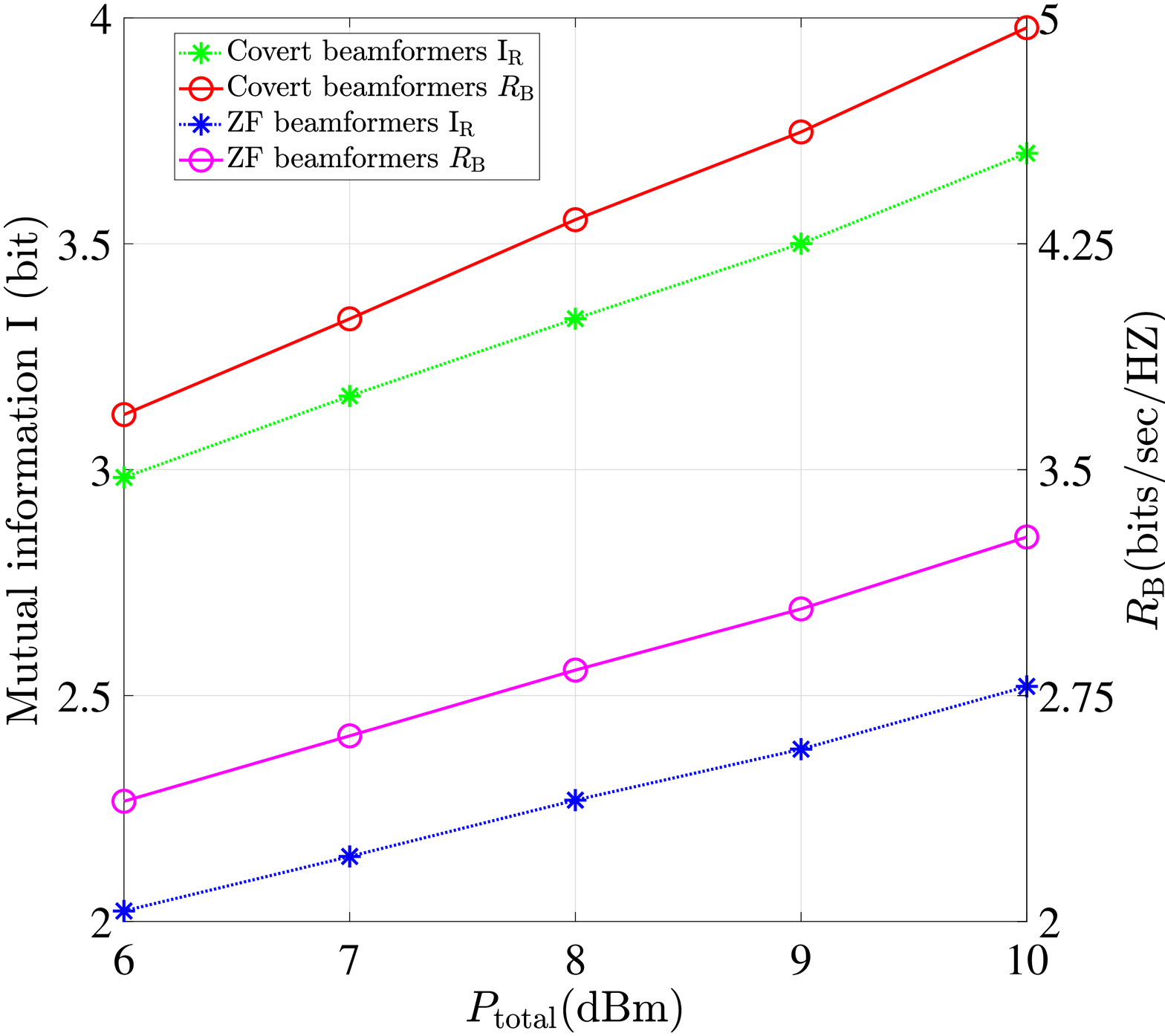}
			\caption{~The proposed covert beamformer design and proposed ZF beamformer design by Mutual information ${\rm{I}} $ (bit) and $R_{\rm{B}}$ (bits/sec/HZ) versus $P_{\rm{total}}$ (dBm), with the number of  antennas $N=5$ }
			\label{P1}
		\end{minipage}
		\begin{minipage}[t]{0.45\textwidth}
			\centering
			\includegraphics[width=7cm,height=6.65cm]{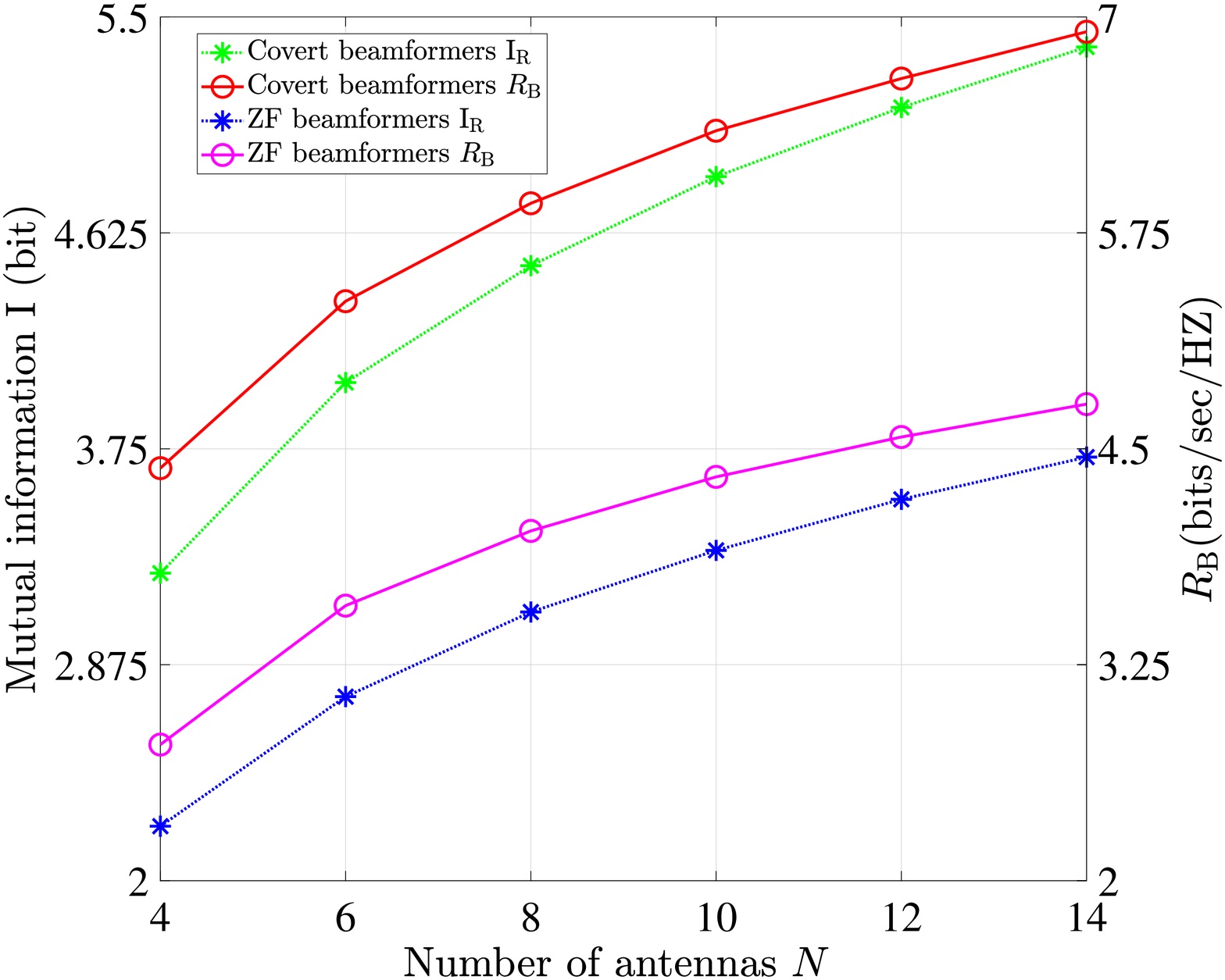}
			\caption{~The proposed covert beamformer design and proposed ZF beamformer design by ~Mutual information ${\rm{I}} $ (bit) and $R_{\rm{B}}$ versus the number of  antennas $N$, with the total transmit power $P_{\rm {total}}= 10\rm{dBm}$ .}
			\label{N1}
		\end{minipage}
	\end{figure}

	Fig. \ref{P1} shows the mutual information $	{\rm{I}}\left( {{{\bf{y}}_{\rm{R}}};{{\bf{h}}_{\rm{T}}}\left| {{s_{\rm{R}}}} \right.} \right){\rm{ }}$ and $R_{\rm{B}}$ with the proposed covert beamformer design and the proposed ZF beamformer design versus the total transmit power $P_{\rm {total}}$, where the number of antennas is set as $N=5$.
	It can be observed that the mutual information of Radar $	{\rm{I}}\left( {{{\bf{y}}_{\rm{R}}};{{\bf{h}}_{\rm{T}}}\left| {{s_{\rm{R}}}} \right.} \right){\rm{ }}$ and $R_{\rm{B}}$ almost linearly increase as
	the total available power $P_{\rm {total}}$ increases.

	In Fig.  \ref{N1}, we plot the mutual information $	{\rm{I}}\left( {{{\bf{y}}_{\rm{R}}};{{\bf{h}}_{\rm{T}}}\left| {{s_{\rm{R}}}} \right.} \right){\rm{ }}$ and $R_{\rm{B}}$ under the proposed covert beamformer design and the proposed ZF beamformer design versus the number of radar antennas $N$, where $P_{\rm {total}}= 10\rm{dBm}$.
	It is observed that  as the number of antennas $N$ increases, the mutual information $	 {\rm{I}}\left( {{{\bf{y}}_{\rm{R}}};{{\bf{h}}_{\rm{T}}}\left| {{s_{\rm{R}}}} \right.} \right){\rm{ }}$ and $R_{\rm{B}}$ increase in a logarithmic fashion, and  the gap between the covert beamformer design and ZF beamformer design   also increases. Because with more  antennas, more   spatial multiplexing gains can be realized. Moreover,  Fig. \ref{P1} and \ref{N1} show that the covert beamforming design can achieve a larger MI and $R_{\rm{B}}$ than those of the   ZF beamformer design.
	
	\subsection{Imperfect WCSI}
	
	\subsubsection{MI Maximization}
	\begin{figure}
		\begin{minipage}[b]{0.45\textwidth}
			\centering
			\includegraphics[width=7cm]{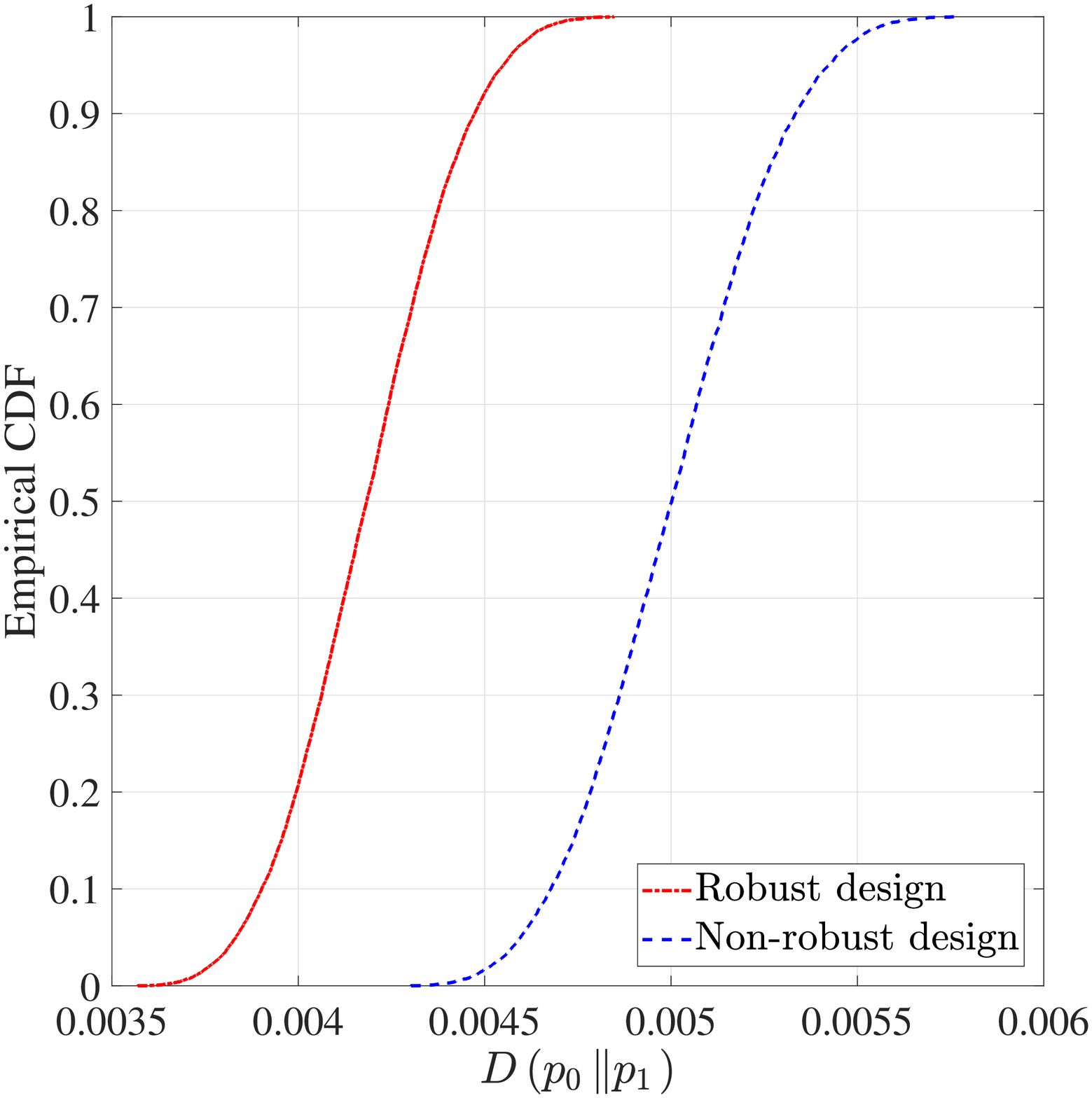}
			\vskip-0.2cm\centering {\footnotesize (a)}
		\end{minipage}\hfill
		\vskip 0.51cm
		\begin{minipage}[b]{0.45\textwidth}
			\centering
			\includegraphics[width=7cm]{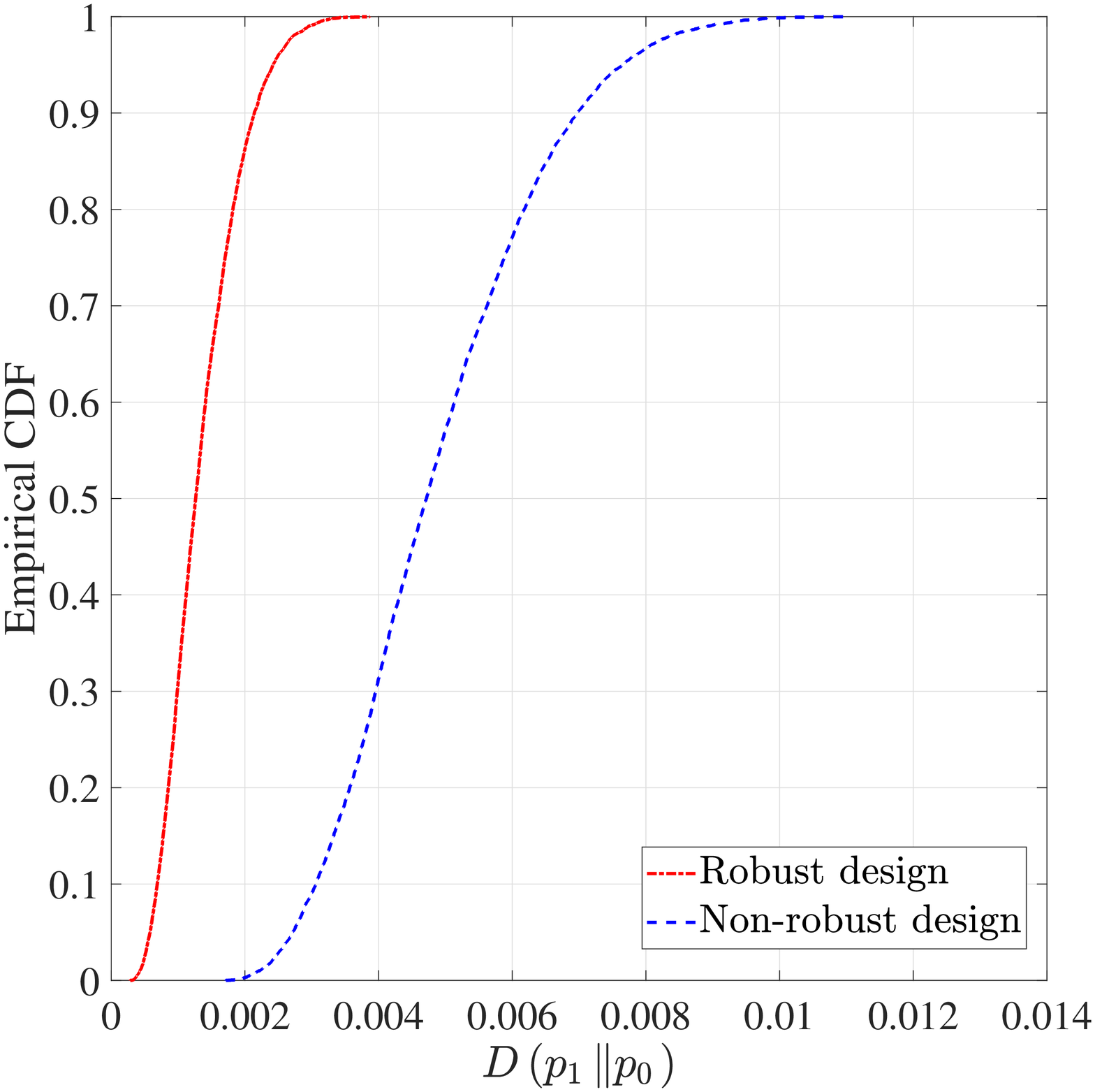}
			\vskip-0.2cm\centering {\footnotesize (b)}
		\end{minipage}\hfill
		\caption{The empirical CDF of (a) $D\left( {{p_0}\left\| {{p_1}} \right.} \right)$  and (b) $D\left( {{p_0}\left\| {{p_1}} \right.} \right)$, with the covertness threshold   $2{\varepsilon ^2} = 0.005$ and  CSI errors   $v_w=0.005$.}
		\label{MI-cdf}  
	\end{figure}
	Fig. \ref{MI-cdf} (a) and (b) show the empirical CDF  of the achieved $D\left( {{p_0}\left\| {{p_1}} \right.} \right)$ and $D\left( {{p_1}\left\| {{p_0}} \right.} \right)$ under the covertness threshold   $2{\varepsilon ^2} = 0.005$ and  CSI errors   $v_w=0.005$, respectively, where $N=5$, and $P_{\rm{total}}=10\rm{dBm}$. Here, the non-robust design refers to the covert design by using the information of ${{{\bf{\hat h}}}_{\rm{W}}} $ only, instead of ${{{\bf{ h}}}_{\rm{W}}} $.
	The covertness thresholds of the robust and non-robust designs are both $2{\varepsilon ^2} = 0.005$, i.e., $D\left( {{p_0}\left\| {{p_1}} \right.} \right) \le 0.005$ and $D\left( {{p_1}\left\| {{p_0}} \right.} \right) \le 0.005$.
	It can be seen that the CDF of the KL divergence of the non-robust design  does not satisfy the constraints, where   about 50\% of the resulting $D\left( {{p_0}\left\| {{p_1}} \right.} \right)$ exceed the covertness threshold
	0.005; and about 57\% of the resulting $D\left( {{p_1}\left\| {{p_0}} \right.} \right)$ exceed the covertness threshold 0.005. On the other hand, the robust beamforming design guarantees the KL divergence requirement, that is,  it satisfies Willie's error detection probability requirement. In general, Fig. \ref{MI-cdf} (a) and (b) demonstrate the effectiveness of the proposed robust design.

	\begin{figure}
		\begin{minipage}[b]{0.45\textwidth}
			\centering
			\includegraphics[width=7cm,height=7.1cm]{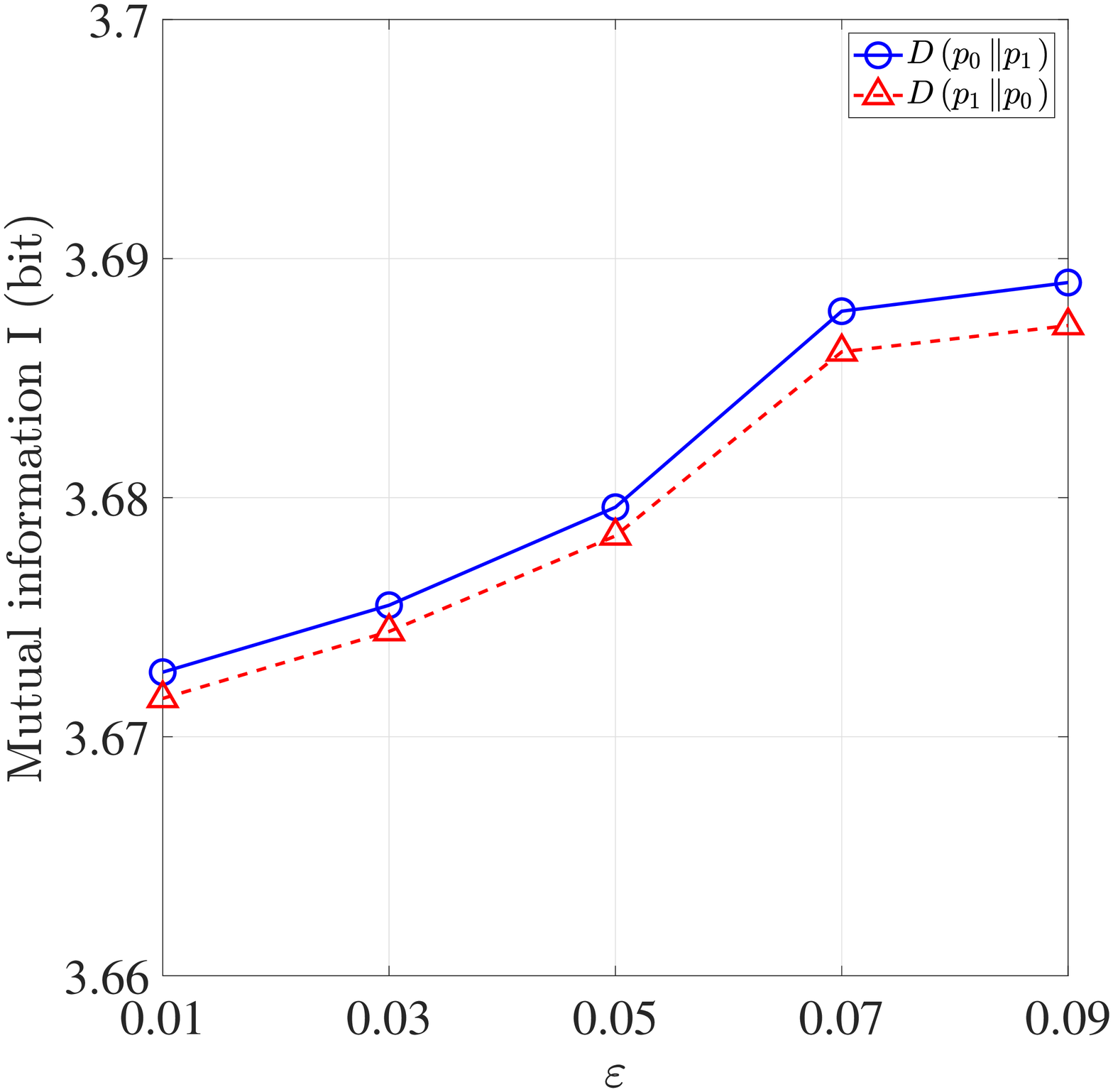}
			\vskip-0.2cm\centering {\footnotesize (a)}
		\end{minipage}\hfill
		\begin{minipage}[b]{0.45\textwidth}
			\centering
			\includegraphics[height=7.45cm]{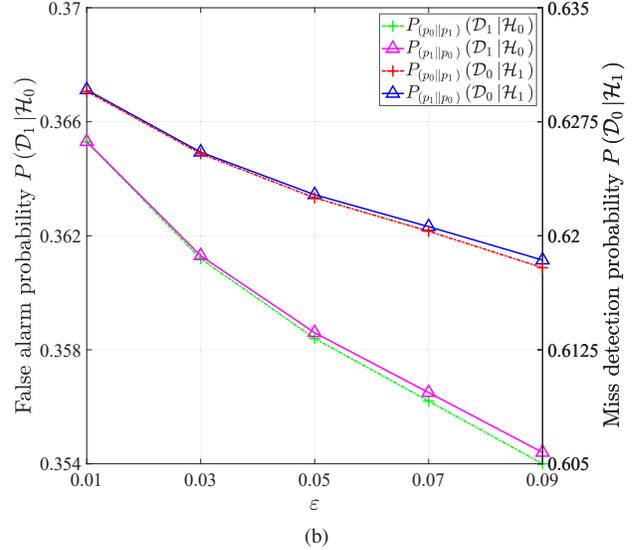}
			\vskip-0.2cm\centering {\footnotesize (b)}
		\end{minipage}\hfill
		\caption{The value of $\varepsilon$ versus (a) the mutual information $\rm{I}$ and (b) the detection error probabilities  with CSI errors   $v_w=0.001$.}
		\label{MI-fig5}  
	\end{figure}

	Fig. \ref{MI-fig5} (a) and (b) show the value of $\varepsilon$ versus the mutual information $	 {\rm{I}}\left( {{{\bf{y}}_{\rm{R}}};{{\bf{h}}_{\rm{T}}}\left| {{s_{\rm{R}}}} \right.} \right){\rm{ }}$ and the detection error probabilities  under the two KL divergence cases, where CSI errors $v_w=0.001$.
	Fig. \ref{MI-fig5} (a) plots   the mutual information $	{\rm{I}}\left( {{{\bf{y}}_{\rm{R}}};{{\bf{h}}_{\rm{T}}}\left| {{s_{\rm{R}}}} \right.} \right){\rm{ }}$ versus the value of $\varepsilon$ under two covertness constraints $D\left( {{p_0}\left\| {{p_1}} \right.} \right) \le 2{\varepsilon ^2}$ and $D\left( {{p_1}\left\| {{p_0}} \right.} \right) \le 2{\varepsilon ^2}$, where CSI errors   $v_w=0.001$. This simulation
	result is consistent with the theoretical analysis, that is, when
	$\varepsilon$ becomes larger, the covertness constraint becomes loose,
	which leads to a larger $	{\rm{I}}\left( {{{\bf{y}}_{\rm{R}}};{{\bf{h}}_{\rm{T}}}\left| {{s_{\rm{R}}}} \right.} \right){\rm{ }}$. On the other hand, $	{\rm{I}}\left( {{{\bf{y}}_{\rm{R}}};{{\bf{h}}_{\rm{T}}}\left| {{s_{\rm{R}}}} \right.} \right){\rm{ }}$ under the covertness constraints $D\left( {{p_0}\left\| {{p_1}} \right.} \right) \le 2{\varepsilon ^2}$ is higher than that
	under the constraint $D\left( {{p_1}\left\| {{p_0}} \right.} \right) \le 2{\varepsilon ^2}$.
	Fig. \ref{MI-fig5} (b) plots the detection error probabilities  for the two KL divergence cases versus $\varepsilon$ , where CSI errors $v_w=0.001$. Here ${P_{\left( {{p_0}\left\| {{p_1}} \right.}\right)}} \left( {{{\cal D}_1}\left| {{{\cal H}_0}} \right.} \right)$ represents the FA probability $P \left( {{{\cal D}_1}\left| {{{\cal H}_0}} \right.} \right)$ in the case of $D\left( {{p_0}\left\| {{p_1}} \right.} \right) \le 2{\varepsilon ^2}$, and the other notation is defined similarly.
	It is found that under the two cases of the covertness constraint, the FA probability $P \left( {{{\cal D}_1}\left| {{{\cal H}_0}} \right.} \right)$ and the MD probability $P \left( {{{\cal D}_0}\left| {{{\cal H}_1}} \right.} \right)$ decrease as    $\varepsilon$ increases, where $P \left( {{{\cal D}_1}\left| {{{\cal H}_0}} \right.} \right)$ is always lower than $P \left( {{{\cal D}_0}\left| {{{\cal H}_1}} \right.} \right)$.
	This shows that the looser the covertness constraint is, the better Willie's detection performance will be. In addition, Fig. \ref{MI-fig5} (b)  also verifies the effectiveness of the proposed robust beamformer design in covert communication.
	
	\begin{figure}
		\begin{minipage}[b]{0.45\textwidth}
			\centering
			\includegraphics[width=7cm]{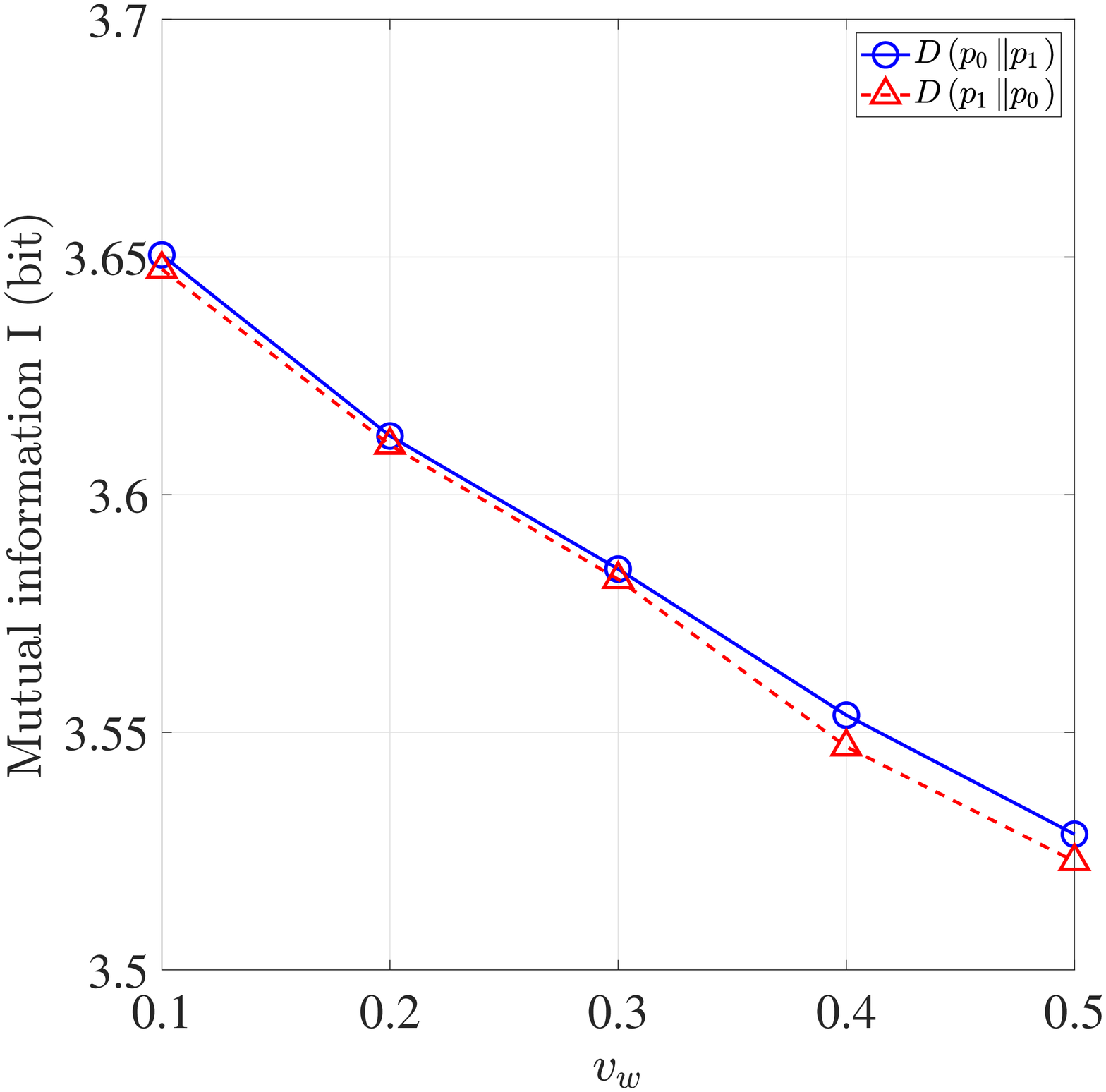}
			\vskip-0.2cm\centering {\footnotesize (a)}
		\end{minipage}
		\begin{minipage}[b]{0.45\textwidth}
			\centering
			\includegraphics[height=7.45cm]{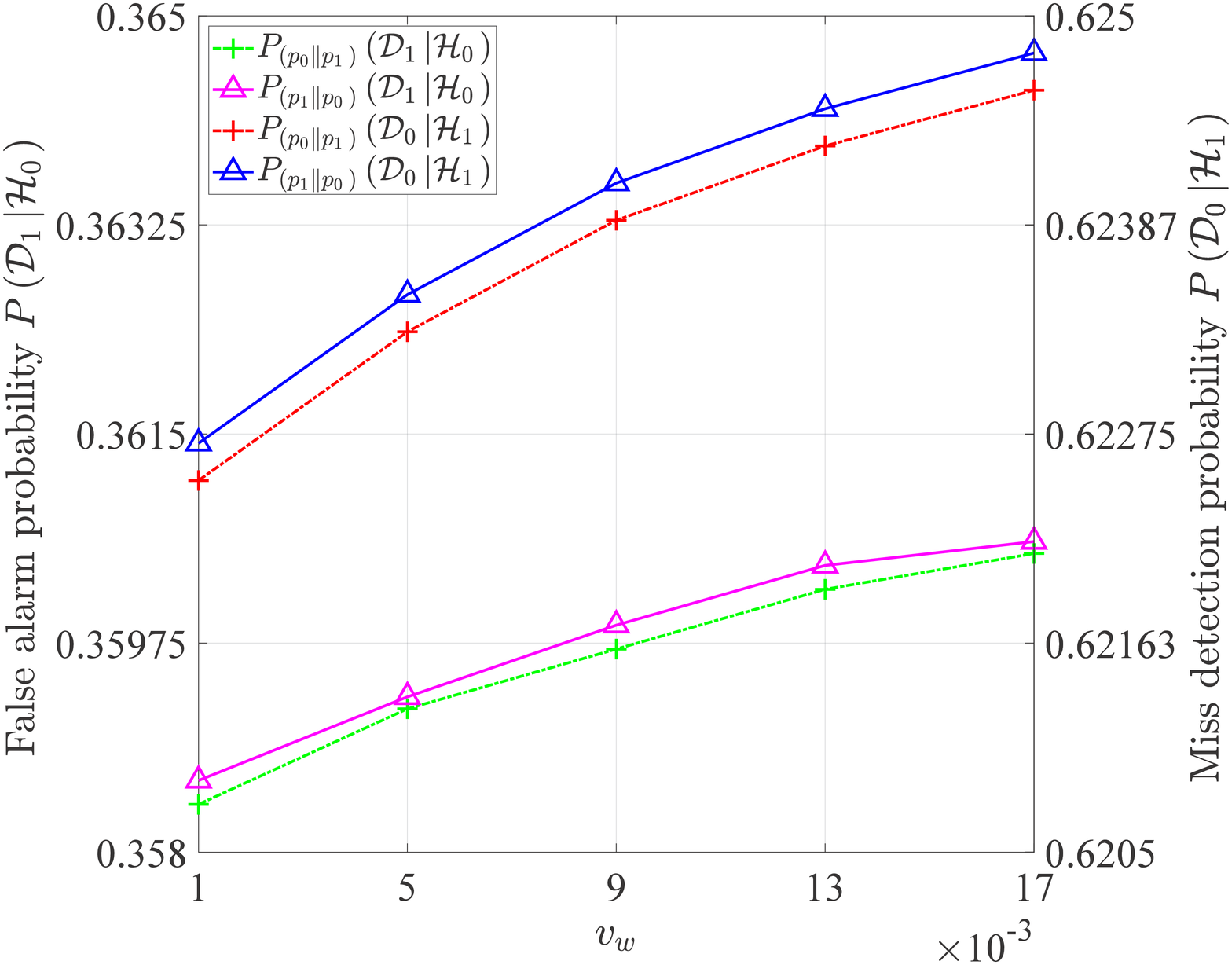}
			\vskip-0.2cm\centering {\footnotesize (b)}
		\end{minipage}\hfill
		\caption{(a) The mutual information $\rm{I}$ versus CSI errors $v_w$ with the value of $\varepsilon=0.01$ and (b) the detection error probabilities
			versus CSI errors $v_w$ with the value of $\varepsilon=0.05$.
		}
		\label{MI-fig6}  
	\end{figure}
	
	Fig. \ref{MI-fig6} (a) and (b) show CSI errors $v_w$ versus the mutual information $	 {\rm{I}}\left( {{{\bf{y}}_{\rm{R}}};{{\bf{h}}_{\rm{T}}}\left| {{s_{\rm{R}}}} \right.} \right){\rm{ }}$ and the detection error probabilities  under
	the two covertness constraints $D\left( {{p_0}\left\| {{p_1}} \right.} \right) \le 2{\varepsilon ^2}$ and $D\left( {{p_1}\left\| {{p_0}} \right.} \right) \le 2{\varepsilon ^2}$, where the value of $\varepsilon=0.01$, $\varepsilon=0.05$, respectively.	
	Fig. \ref{MI-fig6} (a) plots  the mutual information $	{\rm{I}}\left( {{{\bf{y}}_{\rm{R}}};{{\bf{h}}_{\rm{T}}}\left| {{s_{\rm{R}}}} \right.} \right){\rm{ }}$ versus CSI errors $v_w$ for the two KL divergence cases. It can be seen that the higher the CSI error is, the higher the mutual information $	{\rm{I}}\left( {{{\bf{y}}_{\rm{R}}};{{\bf{h}}_{\rm{T}}}\left| {{s_{\rm{R}}}} \right.} \right){\rm{ }}$ and the worse radar performance will be.
	Fig. \ref{MI-fig6} (b) plots the FA probability $P \left( {{{\cal D}_1}\left| {{{\cal H}_0}} \right.} \right)$ and the MD probability $P \left( {{{\cal D}_0}\left| {{{\cal H}_1}} \right.} \right)$ versus  CSI errors $v_w$ under
	two covertness constraints $D\left( {{p_0}\left\| {{p_1}} \right.} \right) \le 2{\varepsilon ^2}$ and $D\left( {{p_1}\left\| {{p_0}} \right.} \right) \le 2{\varepsilon ^2}$, where the value of $\varepsilon=0.05$. We observe that under the two covertness constraints, the FA probability $P \left( {{{\cal D}_1}\left| {{{\cal H}_0}} \right.} \right)$ and the MD probability $P \left( {{{\cal D}_0}\left| {{{\cal H}_1}} \right.} \right)$ both increase with the increase of $v_w$, where $P \left( {{{\cal D}_1}\left| {{{\cal H}_0}} \right.} \right)$ is always less than $P \left( {{{\cal D}_0}\left| {{{\cal H}_1}} \right.} \right)$.
	Moreover, Fig. \ref{MI-fig6} implies that a large error $v_w$  may lead to a good beamformer design in terms  of radar performance and this beamformer may interfere with Willie's detection, which is also beneficial to Bob. The intuition is that when the CSI error is large, the detection performance at Willie clearly would degrade.
	
	\begin{figure}[htbp]
		\begin{minipage}[t]{0.45\textwidth}
			\centering
			\includegraphics[width=7cm]{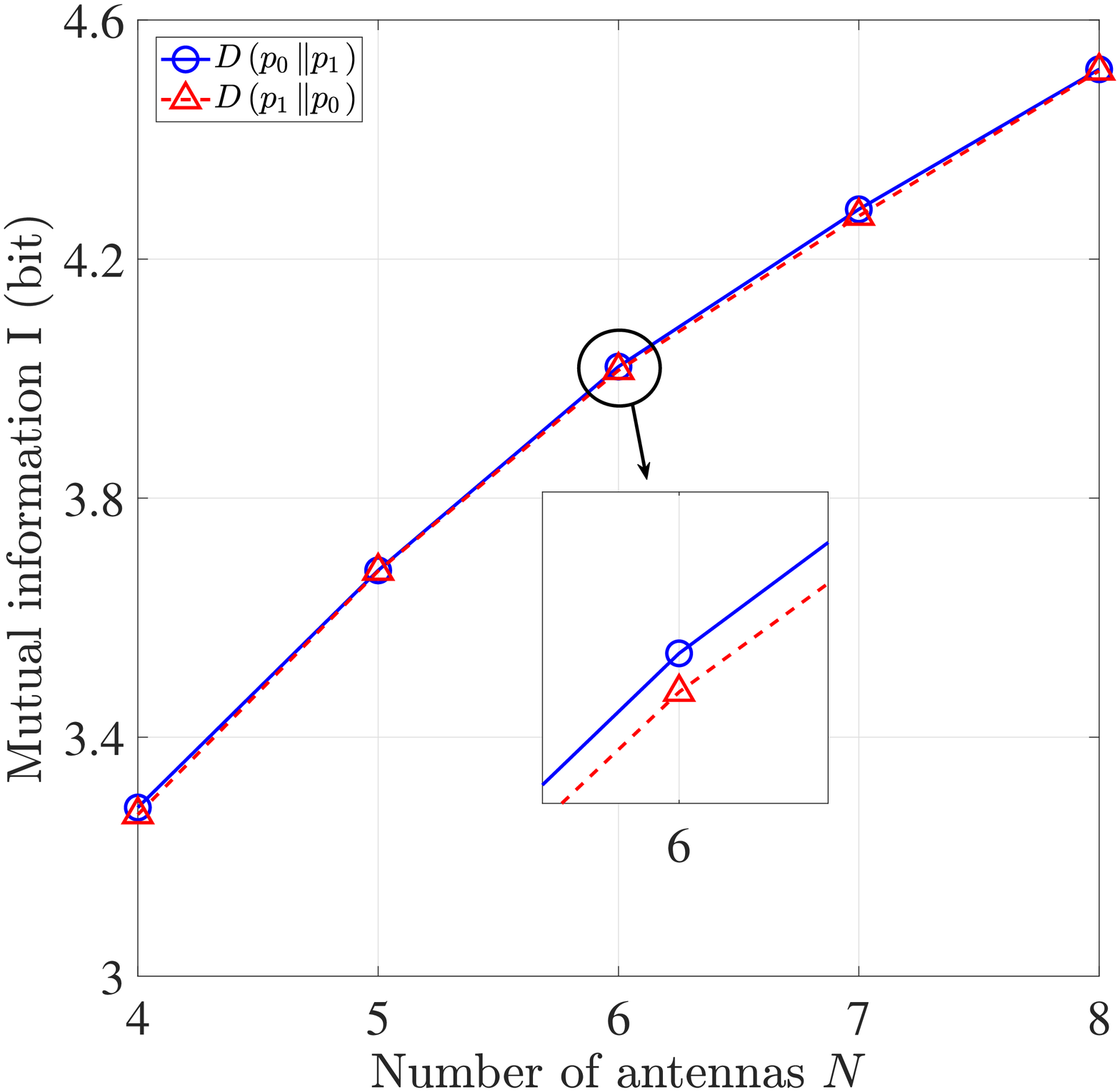}
			\caption{ The mutual information $\rm{I}$ versus  number of  antennas $N$  with CSI errors   $v_w=0.005$, $\varepsilon = 0.05$.}
			\label{MI-2_N} 
		\end{minipage}
		\begin{minipage}[t]{0.45\textwidth}
			\centering
			\includegraphics[width=7cm]{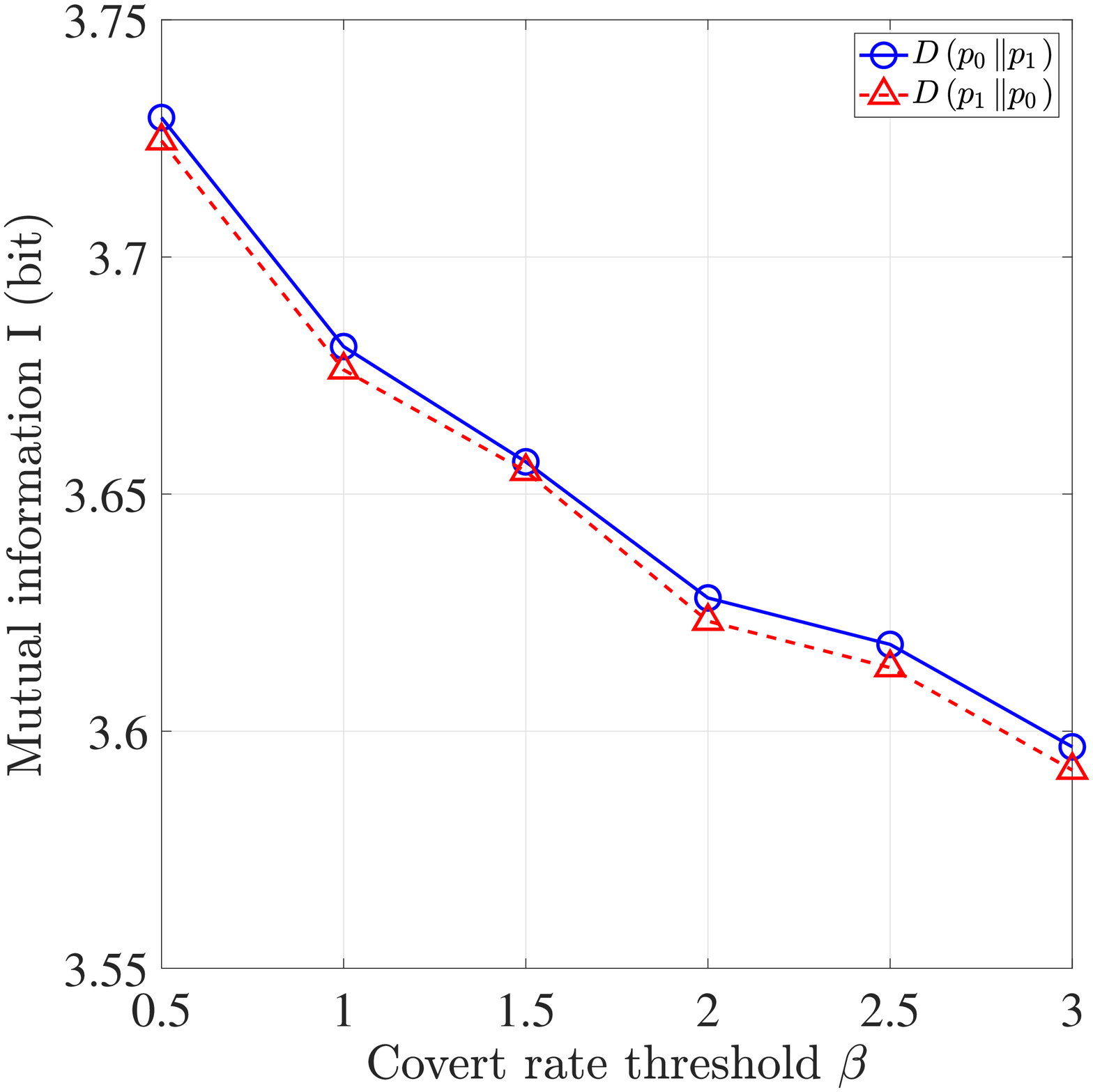}
			\caption{ The mutual information $\rm{I}$ versus covert rates threshold $\beta$  with CSI errors   $v_w=0.005$, $\varepsilon = 0.05$.}
			\label{MI-b}
		\end{minipage}
	\end{figure}
	
	
	Fig. \ref{MI-2_N} shows  the mutual information $	{\rm{I}}\left( {{{\bf{y}}_{\rm{R}}};{{\bf{h}}_{\rm{T}}}\left| {{s_{\rm{R}}}} \right.} \right){\rm{ }}$ versus the number of  antennas $N$ under
	two covertness constraints $D\left( {{p_0}\left\| {{p_1}} \right.} \right) \le 2{\varepsilon ^2}$ and $D\left( {{p_1}\left\| {{p_0}} \right.} \right) \le 2{\varepsilon ^2}$, where  $v_w=0.005$, $\varepsilon = 0.05$.
	From  Fig. \ref{MI-2_N}, we can see that the higher the number of antennas $N$ is, the higher the mutual information $	{\rm{I}}\left( {{{\bf{y}}_{\rm{R}}};{{\bf{h}}_{\rm{T}}}\left| {{s_{\rm{R}}}} \right.} \right){\rm{ }}$ will be, which is similar to the case in Fig. \ref{N1}.
	
	Finally, Fig. \ref{MI-b} shows  the MI $	{\rm{I}}\left( {{{\bf{y}}_{\rm{R}}};{{\bf{h}}_{\rm{T}}}\left| {{s_{\rm{R}}}} \right.} \right){\rm{ }}$ versus  covert rates threshold $\beta$  under
	two covertness constraints $D\left( {{p_0}\left\| {{p_1}} \right.} \right) \le 2{\varepsilon ^2}$ and $D\left( {{p_1}\left\| {{p_0}} \right.} \right) \le 2{\varepsilon ^2}$, where  $v_w=0.001$, $\varepsilon = 0.05$. We can see that as the covert rate threshold $\beta$ increases, the mutual information $	{\rm{I}}\left( {{{\bf{y}}_{\rm{R}}};{{\bf{h}}_{\rm{T}}}\left| {{s_{\rm{R}}}} \right.} \right){\rm{ }}$ gradually decreases.
	Moreover,  from Fig. \ref{MI-fig5}-\ref{MI-b}, we observe  that the mutual information $\rm{I}$ with the covertness constraint $D\left( {{p_0}\left\| {{p_1}} \right.} \right) \le 2{\varepsilon ^2}$
	is higher than that with   the covertness constraint  $D\left( {{p_1}\left\| {{p_0}} \right.} \right) \le 2{\varepsilon ^2}$. This is because
	$D\left( {{p_1}\left\| {{p_0}} \right.} \right) \le 2{\varepsilon ^2}$ is stricter than
	$D\left( {{p_0}\left\| {{p_1}} \right.} \right) \le 2{\varepsilon ^2}$, and  this conclusion is in line with \cite{Yan_TWC_2019}.
	\subsubsection{Rate Maximization}
	In this subsection, we set $N=5$ and $P_{\rm{total}}=10\rm{dBm}$.
	\begin{figure}
		\begin{minipage}[b]{0.45\textwidth}
			\centering
			\includegraphics[width=7cm]{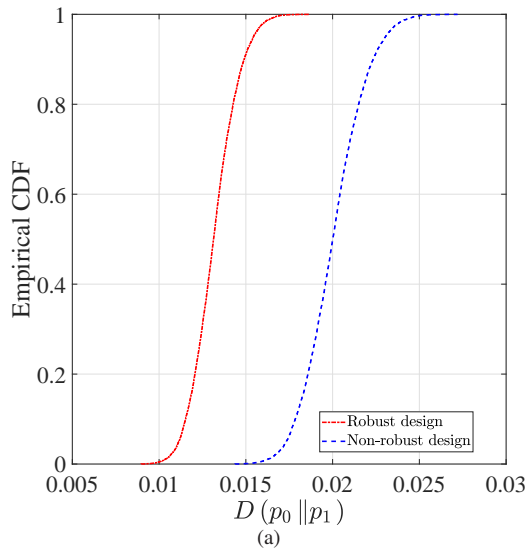}
			\vskip-0.2cm\centering {\footnotesize (a)}
		\end{minipage}\hfill
		\vskip 0.4cm
		\begin{minipage}[b]{0.45\textwidth}
			\centering
			\includegraphics[width=7cm]{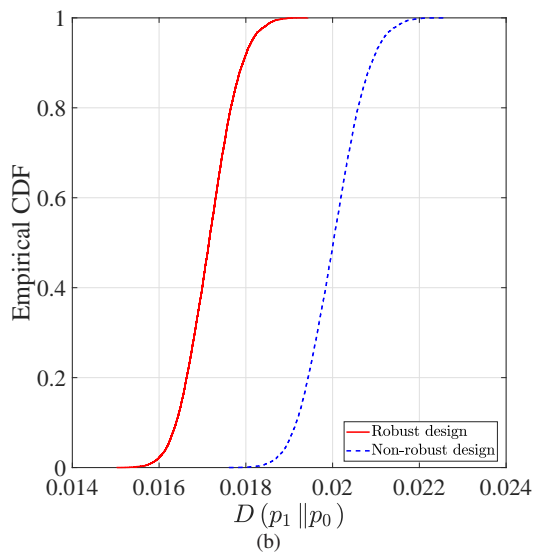}
			\vskip-0.2cm\centering {\footnotesize (b)}
		\end{minipage}\hfill
		\caption{The empirical CDF of (a) $D\left( {{p_0}\left\| {{p_1}} \right.} \right)$  and (b) $D\left( {{p_0}\left\| {{p_1}} \right.} \right)$, with the covertness threshold   $2{\varepsilon ^2} = 0.02$ and  CSI errors   $v_w=0.001$.}
		\label{cdf}  
	\end{figure}
	
	Fig. \ref{cdf} (a) and (b) show the empirical CDF  of the achieved $D\left( {{p_0}\left\| {{p_1}} \right.} \right)$ and $D\left( {{p_1}\left\| {{p_0}} \right.} \right)$ under the covertness threshold   $2{\varepsilon ^2} = 0.02$ and  CWSI errors   $v_w=0.001$, respectively. The covertness threshold of the robust and non-robust designs are both $2{\varepsilon ^2} = 0.02$, i.e., $D\left( {{p_0}\left\| {{p_1}} \right.} \right) \le 0.02$ and $D\left( {{p_1}\left\| {{p_0}} \right.} \right) \le 0.02$.
	It can be seen that the CDF of the KL divergence of the non-robust design cannot satisfy the constraints. On the other hand, the robust beamforming design guarantees the KL divergence requirement, that is,  it satisfies Willie's error
	detection probability requirement.  Here,
	the non-robust design refers to the proposed covert design
	with ${{{\bf{\hat h}}}_{\rm{W}}} $ under the same conditions.
	In general, Fig. \ref{cdf} (a) and (b) demonstrate the effectiveness of the proposed robust design.

	\begin{figure}
		\begin{minipage}[b]{0.45\textwidth}
			\centering
			\includegraphics[width=7cm]{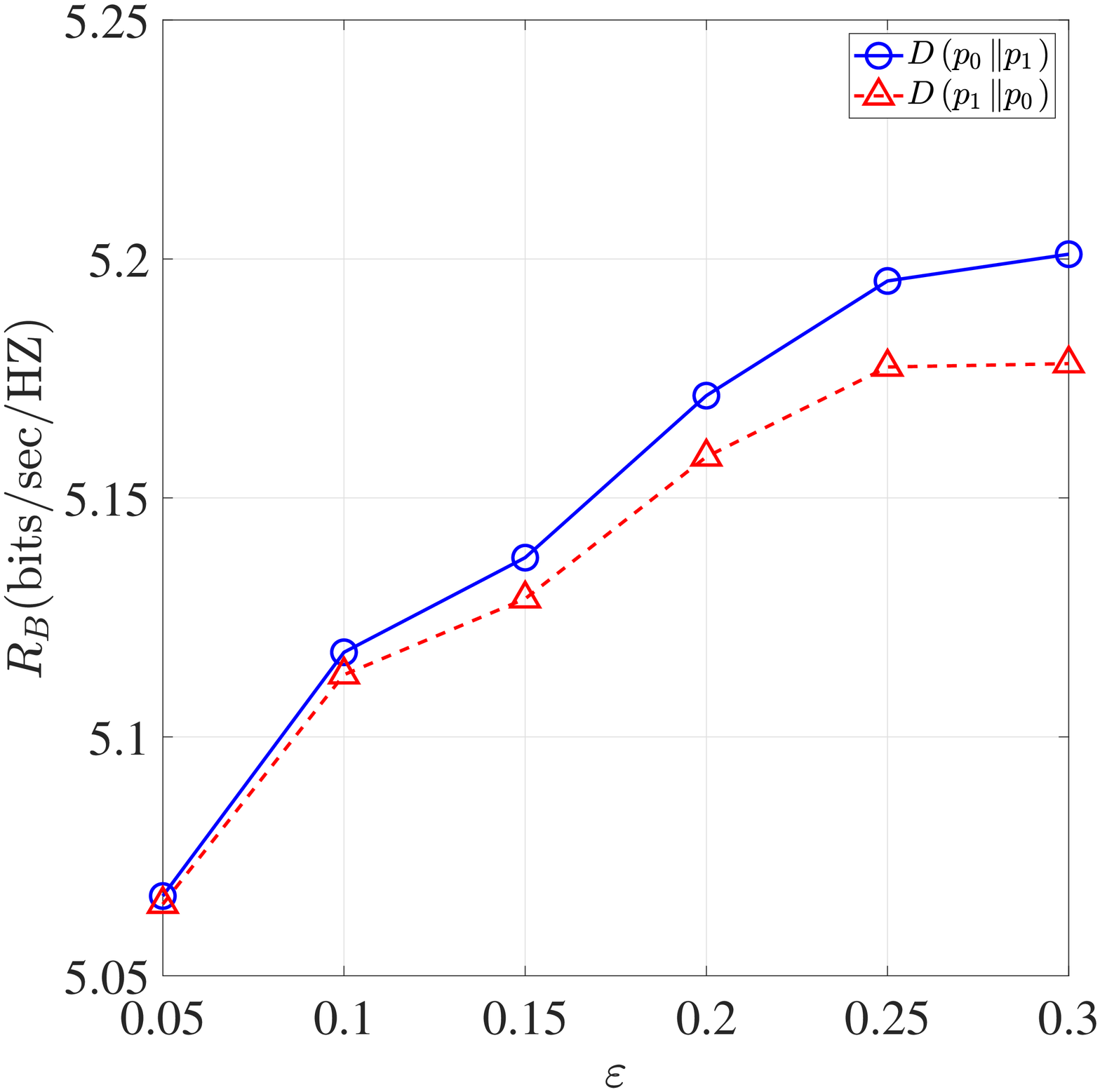}
			\vskip-0.2cm\centering {\footnotesize (a)}
		\end{minipage}
		\begin{minipage}[b]{0.45\textwidth}
			\centering
			\includegraphics[height=7.4cm]{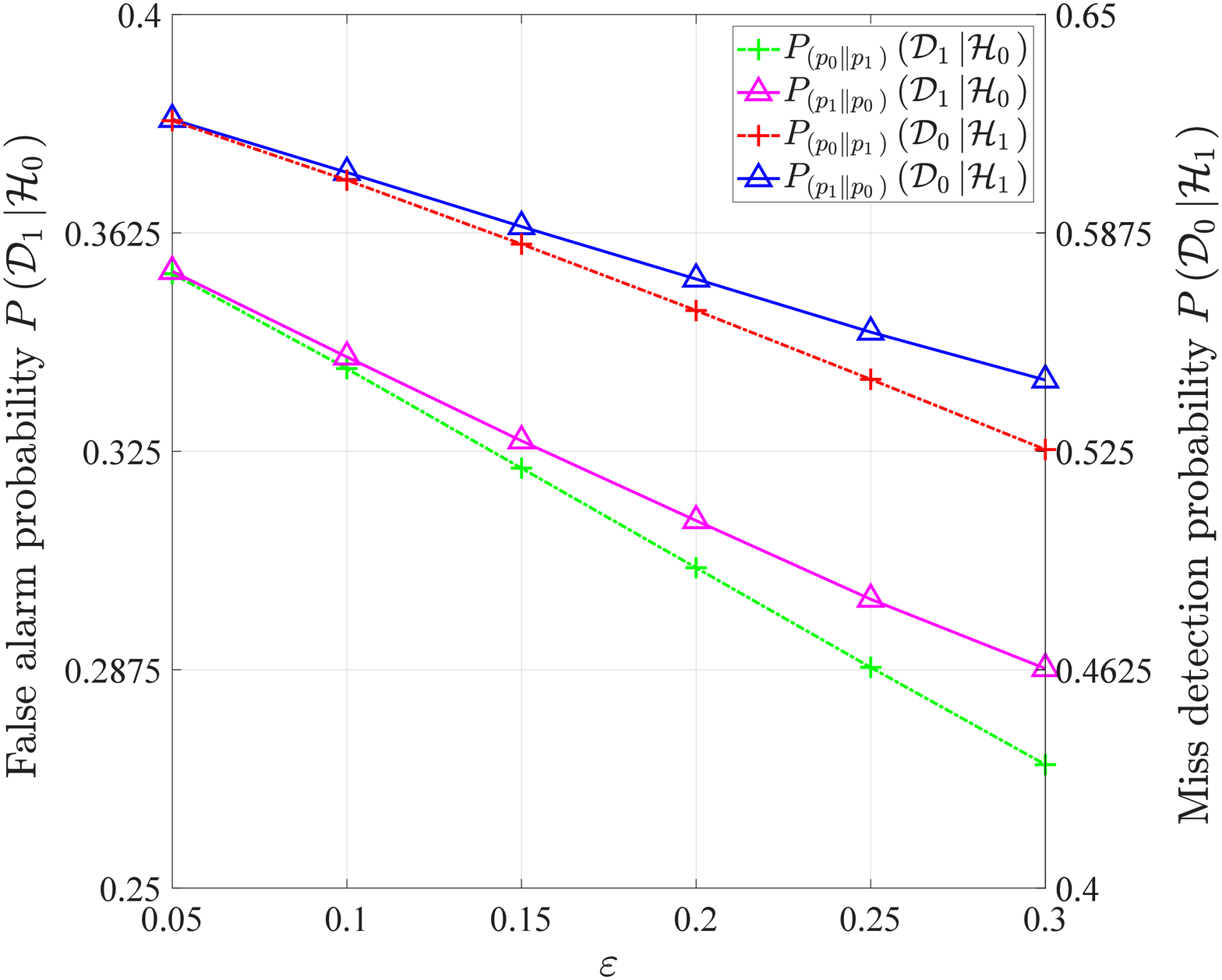}
			\vskip-0.2cm\centering {\footnotesize (b)}
		\end{minipage}\hfill
		\caption{The value of $\varepsilon$ versus (a) the covert rate and (b) the detection error probabilities  with CSI errors   $v_w=0.001$.}
		\label{fig5}  
	\end{figure}

	Fig. \ref{fig5} (a) and (b) show the value of $\varepsilon$ versus the covert rate $R_{\rm{B}}$ and the detection error probabilities  under the two KL divergence cases, where CSI errors $v_w=0.001$.
	Fig. \ref{fig5} (a) plots covert rates $R_{\rm{B}}$ versus $\varepsilon$ under two covertness constraints $D\left( {{p_0}\left\| {{p_1}} \right.} \right) \le 2{\varepsilon ^2}$ and $D\left( {{p_1}\left\| {{p_0}} \right.} \right) \le 2{\varepsilon ^2}$, where CSI errors $v_w=0.001$. This simulation
	result verifies the theoretical analysis that when
	$\varepsilon$ becomes larger, the covertness constraint becomes loose and $R_{\rm{B}}$ becomes larger. On the other hand, $R_{\rm{B}}$ under the covertness constraints $D\left( {{p_0}\left\| {{p_1}} \right.} \right) \le 2{\varepsilon ^2}$ is higher than that under $D\left( {{p_1}\left\| {{p_0}} \right.} \right) \le 2{\varepsilon ^2}$, which shows that  $D\left( {{p_1}\left\| {{p_0}} \right.} \right) \le 2{\varepsilon ^2}$ is a stricter covertness
	constraint than  $D\left( {{p_0}\left\| {{p_1}} \right.} \right) \le 2{\varepsilon ^2}$.
	Fig. \ref{fig5} (b) plots the detection error probabilities for the two KL divergence cases versus $\varepsilon$, where CSI errors $v_w=0.001$. Here ${P_{\left( {{p_0}\left\| {{p_1}} \right.}\right)}} \left( {{{\cal D}_1}\left| {{{\cal H}_0}} \right.} \right)$ represents the FA probability $P \left( {{{\cal D}_1}\left| {{{\cal H}_0}} \right.} \right)$ in the case of $D\left( {{p_0}\left\| {{p_1}} \right.} \right) \le 2{\varepsilon ^2}$, and the other notation is defined similarly.
	It is found that under the two cases of the covertness constraint, the FA probability $P \left( {{{\cal D}_1}\left| {{{\cal H}_0}} \right.} \right)$ and the MD probability $P \left( {{{\cal D}_0}\left| {{{\cal H}_1}} \right.} \right)$ decrease as    $\varepsilon$ increases, where $P \left( {{{\cal D}_1}\left| {{{\cal H}_0}} \right.} \right)$ is always lower than $P \left( {{{\cal D}_0}\left| {{{\cal H}_1}} \right.} \right)$.
	This shows that the looser the covertness constraint is, the better Willie's detection performance will be. In addition, Fig. \ref{fig5} (b)  also verifies the effectiveness of the proposed robust beamformer design in covert communication.
	
	\begin{figure}
		\begin{minipage}[b]{0.45\textwidth}
			\centering
			\includegraphics[width=7cm]{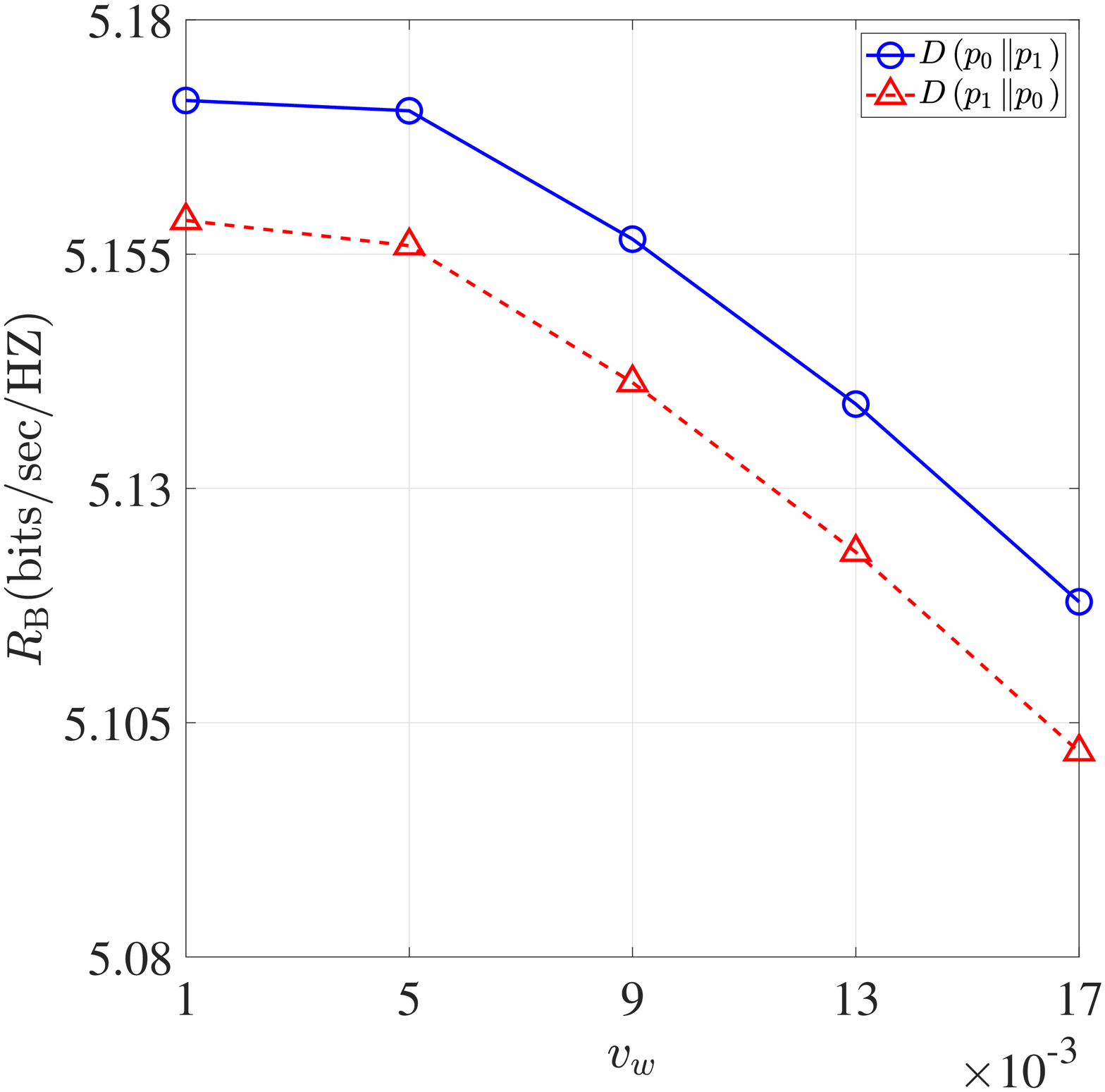}
			\vskip-0.2cm\centering {\footnotesize (a)}
		\end{minipage}\hfill
		\vskip 0.1cm
		\begin{minipage}[b]{0.45\textwidth}
			\centering
			\includegraphics[height=7.4cm]{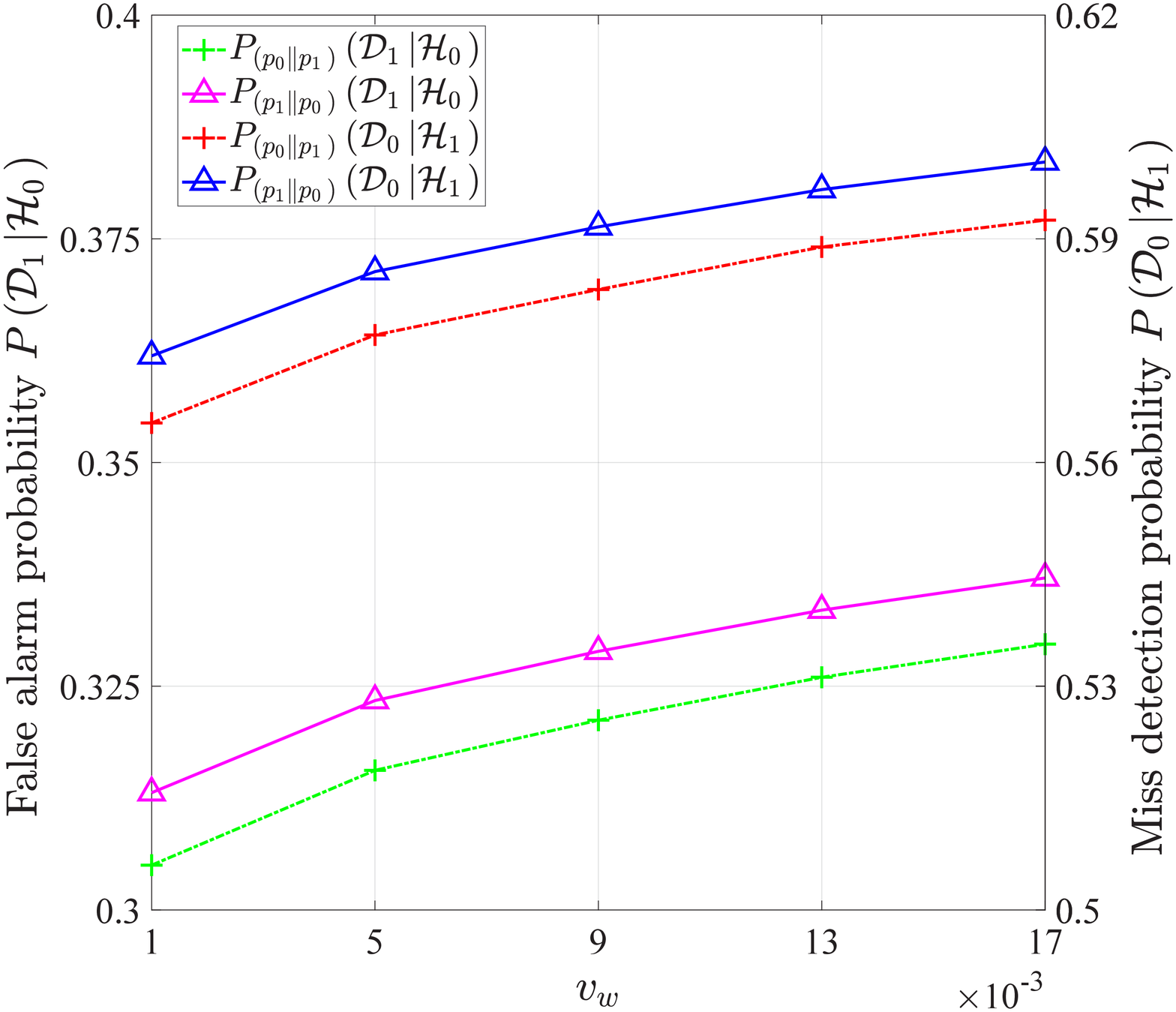}
			\vskip-0.2cm\centering {\footnotesize (b)}
		\end{minipage}\hfill
		\caption{(a) The covert rate and (b) the detection error probabilities
			versus CSI errors $v_w$ with the value of $\varepsilon=0.20$.
		}
		\label{fig6}  
	\end{figure}
	
	Fig. \ref{fig6} (a) and (b) show CSI errors $v_w$ versus the covert rate $R_{\rm{B}}$ and the detection error probabilities  under
	the two covertness constraints $D\left( {{p_0}\left\| {{p_1}} \right.} \right) \le 2{\varepsilon ^2}$ and $D\left( {{p_1}\left\| {{p_0}} \right.} \right) \le 2{\varepsilon ^2}$, where the value of $\varepsilon=0.20$.
	Fig. \ref{fig6} (a) plots   covert rates $R_{\rm{B}}$ versus CSI errors $v_w$ for the two KL divergence cases. It can be seen that the higher the CSI error is, the lower the covert rate $R_{\rm{B}}$ will be.
	Fig. \ref{fig6} (b) plots the FA probability $P \left( {{{\cal D}_1}\left| {{{\cal H}_0}} \right.} \right)$ and the MD probability $P \left( {{{\cal D}_0}\left| {{{\cal H}_1}} \right.} \right)$ versus  CSI errors $v_w$ under the
	two covertness constraints $D\left( {{p_0}\left\| {{p_1}} \right.} \right) \le 2{\varepsilon ^2}$ and $D\left( {{p_1}\left\| {{p_0}} \right.} \right) \le 2{\varepsilon ^2}$, where the value of $\varepsilon=0.05$. We observe that under the two covertness constraints, the FA probability $P \left( {{{\cal D}_1}\left| {{{\cal H}_0}} \right.} \right)$ and the MD probability $P \left( {{{\cal D}_0}\left| {{{\cal H}_1}} \right.} \right)$ both increase with the increase of $v_w$, where $P \left( {{{\cal D}_1}\left| {{{\cal H}_0}} \right.} \right)$ is always less than $P \left( {{{\cal D}_0}\left| {{{\cal H}_1}} \right.} \right)$. Similar to the case of MI maximization, a large error $v_w$ may lead to a worse detection performance for Willie.
	\begin{figure}[htbp]
		\begin{minipage}[t]{0.45\textwidth}
			\centering
			\includegraphics[width=7cm]{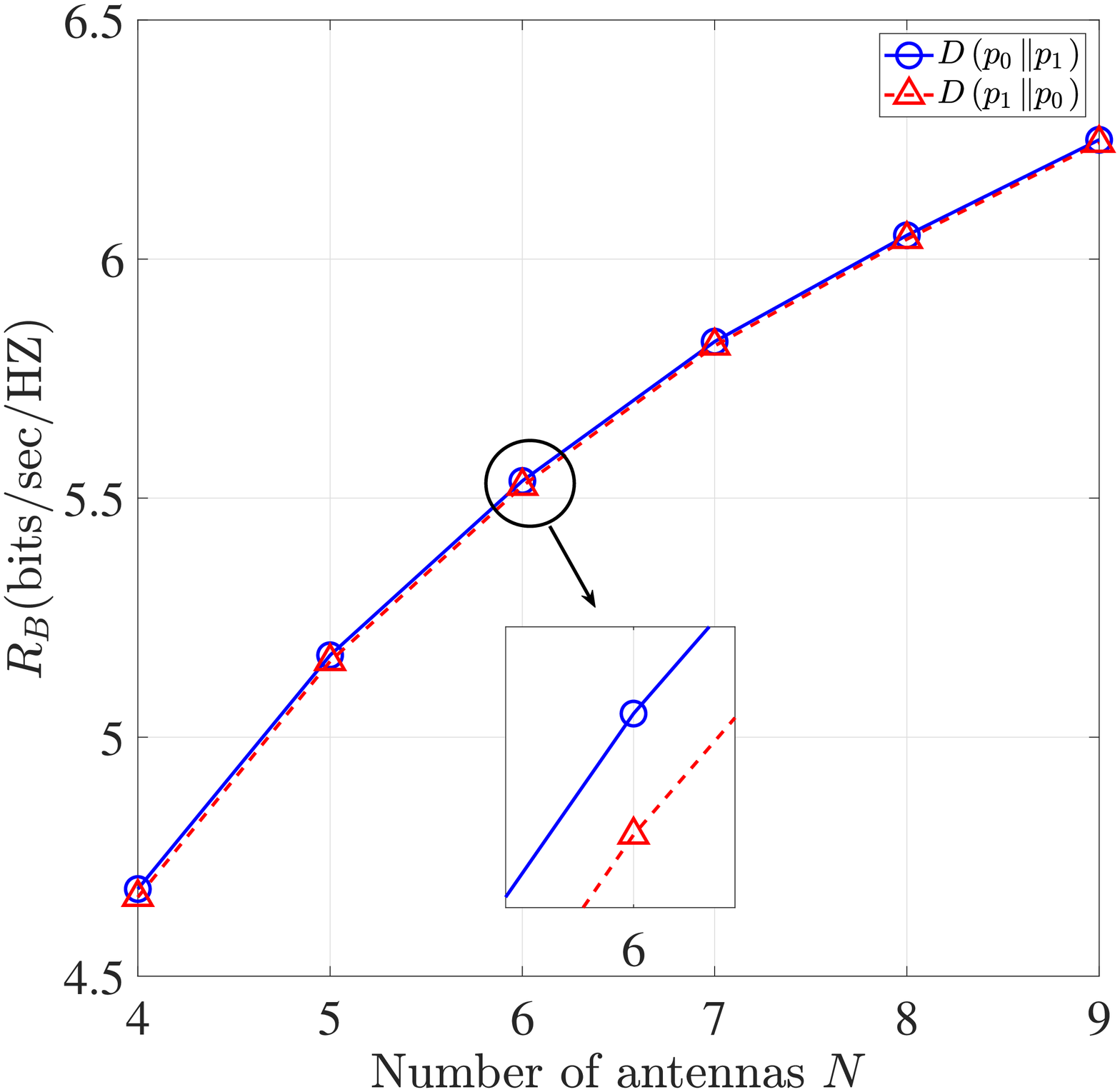}
			\caption{ Covert rates $R_{\rm{B}}$ versus  number of  antennas $N$  with CSI errors   $v_w=0.001$, $\varepsilon=0.20$.}
			\label{2_N_rb}
		\end{minipage}
		\begin{minipage}[t]{0.45\textwidth}
			\centering
			\includegraphics[width=7cm]{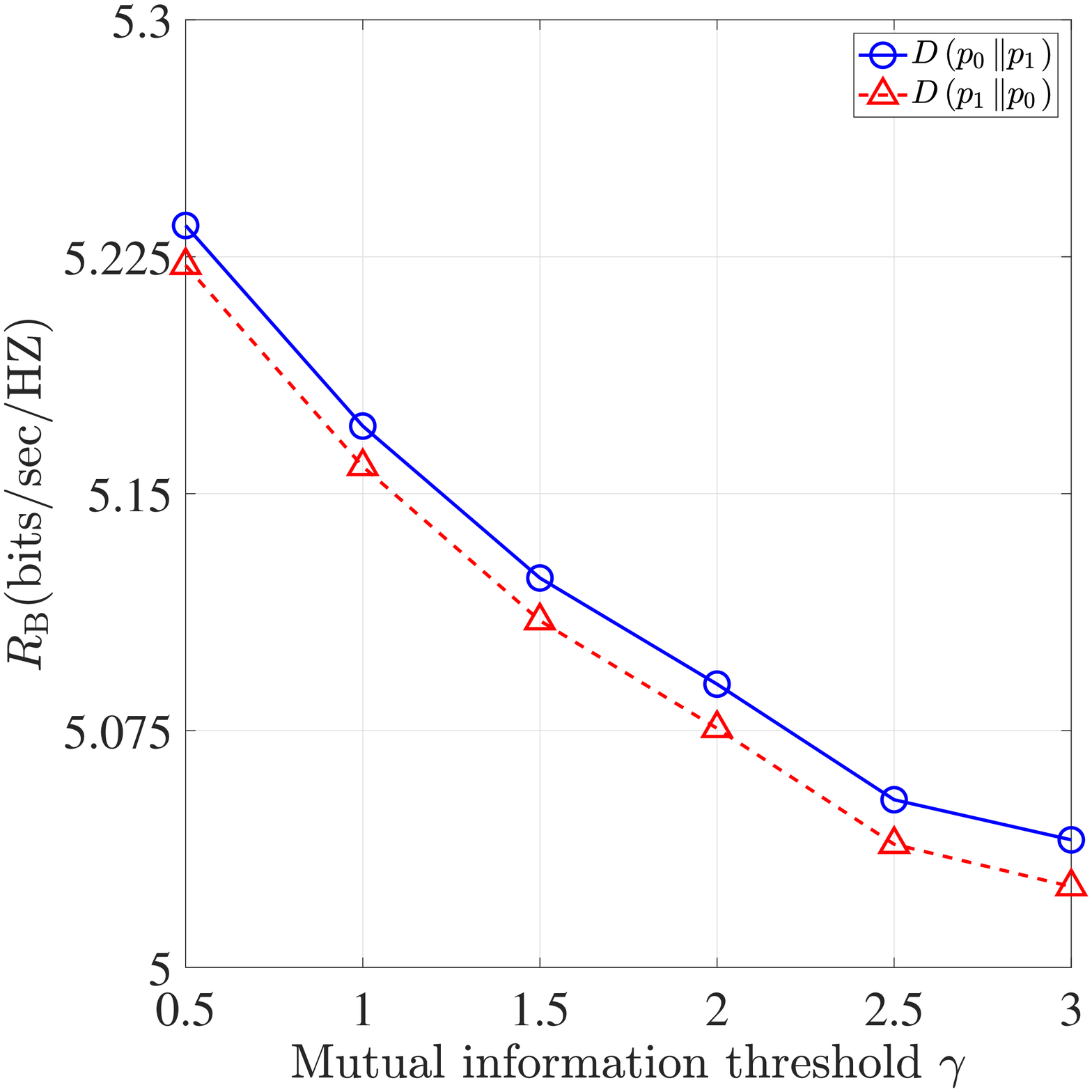}
			\caption{ Covert rates $R_{\rm{B}}$ versus  mutual information threshold $\gamma$  with CSI errors   $v_w=0.001$, $\varepsilon = 0.20$.}
			\label{b}
		\end{minipage}
	\end{figure}
	
	
	Fig. \ref{2_N_rb} shows  the covert rate $R_{\rm{B}}$ versus the number of  antennas $N$ under the
	two covertness constraints $D\left( {{p_0}\left\| {{p_1}} \right.} \right) \le 2{\varepsilon ^2}$ and $D\left( {{p_1}\left\| {{p_0}} \right.} \right) \le 2{\varepsilon ^2}$, where  $v_w=0.001$ and $\varepsilon=0.20$.
	From  Fig. \ref{2_N_rb}, we can see that the higher the number of antennas $N$ is, the higher the achieved covert rates $R_{\rm{B}}$ will be, which is similar to the case in Fig. \ref{N1}.
	Finally, Fig. \ref{b} shows  the covert rates $R_{\rm{B}}$ versus  the mutual information threshold $\gamma$  under the
	two covertness constraints $D\left( {{p_0}\left\| {{p_1}} \right.} \right) \le 2{\varepsilon ^2}$ and $D\left( {{p_1}\left\| {{p_0}} \right.} \right) \le 2{\varepsilon ^2}$, where  $v_w=0.001$, $\varepsilon = 0.2$. We can see that as the mutual information threshold $\gamma$ increases, the covert rates $R_{\rm{B}}$ gradually decreases.
	Moreover,  From Fig. \ref{fig5}-\ref{b}, we observe that the rates with the covertness constraint $D\left( {{p_0}\left\| {{p_1}} \right.} \right) \le 2{\varepsilon ^2}$
	are higher than those with   the covertness constraint  $D\left( {{p_1}\left\| {{p_0}} \right.} \right) \le 2{\varepsilon ^2}$. This is because
	$D\left( {{p_1}\left\| {{p_0}} \right.} \right) \le 2{\varepsilon ^2}$ is stricter than
	$D\left( {{p_0}\left\| {{p_1}} \right.} \right) \le 2{\varepsilon ^2}$, and  this conclusion is also verified in \cite{Yan_TWC_2019}.
	\section{Conclusions}
	In this paper, we developed a covert beamforming design framework for
	IRSC systems. Specifically, we proposed two effective solutions to maximize the MI of the radar while meeting the covertness requirements and the total power constraints to ensure covert transmission. When the WCSI is accurately known, we proposed a single-iterative beamforming design method based on the zero forcing criterion. When the WCSI can only be estimated with error, we proposed a robust optimization method to ensure the worst-case covert IRSC performance. Finally, simulation results
	showed that the proposed covert beamforming design framework  can  simultaneously realize radar detection and covert communication for perfect and imperfect WCSI scenarios.

	 %
	
	%
	%
	\bibliographystyle{IEEE-unsorted}
	\bibliography{refs0611}

\end{document}